\documentclass[preprint,12pt]{elsarticle}

\usepackage{amsfonts}
\usepackage{amsmath}
\usepackage{amssymb}
\usepackage{graphicx}
\usepackage{hyphenat}

\usepackage[dvipdfm]{hyperref}

\usepackage{changes}
\usepackage{color}
\def\erf{\mbox{erf}}

\newcommand{\red}{\color{black}}

\newcommand{\black}{\color{black}}

\journal{Digital Signal Processing}

\begin{document}

\begin{frontmatter}

%% Title, authors and addresses

%% use the tnoteref command within \title for footnotes;
%% use the tnotetext command for theassociated footnote;
%% use the fnref command within \author or \address for footnotes;
%% use the fntext command for theassociated footnote;
%% use the corref command within \author for corresponding author footnotes;
%% use the cortext command for theassociated footnote;
%% use the ead command for the email address,
%% and the form \ead[url] for the home page:
%% \title{Title\tnoteref{label1}}
%% \tnotetext[label1]{}
%% \author{Name\corref{cor1}\fnref{label2}}
%% \ead{email address}
%% \ead[url]{home page}
%% \fntext[label2]{}
%% \cortext[cor1]{}
%% \address{Address\fnref{label3}}
%% \fntext[label3]{}

\title{Bayesian Evidence and Model Selection}

%% use optional labels to link authors explicitly to addresses:
%% \author[label1,label2]{}
%% \address[label1]{}
%% \address[label2]{}

\author{Kevin H. Knuth{$^{1,2*}$}, Michael Habeck{$^{3,4}$}, Nabin K. Malakar{$^{5}$}, Asim M. Mubeen{$^{1,6}$}, Ben Placek{$^{1}$}}

\address{
1. Dept. of Physics, Univ. at Albany (SUNY), Albany NY 12222, USA \\
2. Dept. of Informatics, Univ. at Albany (SUNY), Albany NY 12222, USA \\
3. Max Planck Institute for Biophysical Chemistry, 37077 G{\"o}ttingen, Germany\\
4. Felix Bernstein Institute for Mathematical Statistics in the Biosciences, University of G{\"o}ttingen, 37077 G{\"o}ttingen, Germany\\
5. Jet Propulsion Laboratory, California Institute of Technology, Pasadena CA 91109, USA\\
6. Geriatrics Division, Nathan Kline Institute, Orangeburg NY 10962, USA\\
}

\begin{abstract}
In this paper we review the concepts of Bayesian evidence and Bayes factors, also known as log odds ratios, and their application to model selection.  The theory is presented along with a discussion of analytic, approximate and numerical techniques.  Specific attention is paid to the Laplace approximation, variational Bayes, importance sampling, thermodynamic integration, and nested sampling and its recent variants.  Analogies to statistical physics, from which many of these techniques originate, are discussed in order to provide readers with deeper insights that may lead to new techniques.  The utility of Bayesian model testing in the domain sciences is demonstrated by presenting four specific practical examples considered within the context of signal processing in the areas of signal detection, sensor characterization, scientific model selection and molecular force characterization.
\end{abstract}

\begin{keyword}
\sep Bayesian signal processing
\sep Bayesian evidence
\sep Model testing
\sep Nested sampling
\sep Odds ratio
%% keywords here, in the form: keyword \sep keyword

%% PACS codes here, in the form: \PACS code \sep code

%% MSC codes here, in the form: \MSC code \sep code
%% or \MSC[2008] code \sep code (2000 is the default)

\end{keyword}

\end{frontmatter}

%% \linenumbers

%% main text

\section{Introduction}
The application of model-based reasoning techniques employing Bayesian probability theory has recently found wide use in signal processing, and in the physical sciences in general \cite{Jaynes:Book}\cite{Gregory:2005}\cite{Candy:2009}\cite{vonToussaint:2011}\cite{Gelman+etal:2013}\cite{vonderLinden:2014}.  In such an approach, it is critical to be able to
statistically compare the probability of one model to another.
This is performed by computing the Bayesian evidence of the two models and comparing them by forming a ratio, which is often referred to as a Bayes factor or the odds ratio.

In this paper, we present an overview of the theory behind Bayesian evidence, discuss various methods of computation, and demonstrate the application in four practical examples of current interest more closely related to signal processing.  We do not aim to cover all of the techniques and applications, as there exists a great number of excellent treatments spanning several decades \cite{Jeffreys:1961}\cite{DeGroot:2004}\cite{Kass+Raftery:1995}\cite{Jaynes:Book}\cite{Zellner+Siow:1980}\cite{Sivia+etal:1993}\cite{Andrieu+etal:1998}\cite{Fitzgerald:2001}\cite{Gregory:2005}\cite{Bernardo+Smith:2009}\cite{Gelman+etal:2013}\cite{MacKay:1995}\cite{MacKay:2003}\cite{Sivia&Skilling}\cite{vonToussaint:2011} as well as a wide variety of applications spread across a great number of fields, such as acoustics \cite{Xiang+Goggans:2003}\cite{Jasa+Xiang:2012}\cite{Goggans+Henderson+Xiang:2013}, astronomy, astrophysics and cosmology \cite{Loredo+Lamb:2002}\cite{Liddle+etal:2006}\cite{Clark+etal:2007}\cite{Heavens:2007}\cite{Liddle:2007}\cite{Jullo+etal:2007}\cite{Trotta:2008}\cite{Liddle:2009}\cite{Feroz+etal:2011}\cite{Feroz:2013}\cite{Debono:2014}, chemistry \cite{Sivia+Carlile:1992}, computer science and machine learning \cite{Beal+Ghahramani:2003}\cite{Marshall+etal:2006}, neural networks \cite{Mackay:1992}\cite{Wolpert:1993}\cite{Penny+Roberts:1999}, neuroscience \cite{Lewicki:1994}\cite{Trujillo+etal:2004}\cite{Woolrich+etal:2009}\cite{Friston+Penny:2011}, nuclear and particle physics \cite{GulamRazul+Fitzgerald+Andrieu:2003}\cite{DeCruz+etal:2011}\cite{Bergstrom:2012}, signal processing \cite{Nallanathan+Fitzgerald:1994}\cite{Roberts:1998}\cite{Kannan+etal:2000}\cite{Roberts+Everson:2001}\cite{Andrieu+Djuric+Doucet:2001}\cite{Penny+Roberts:2002}\cite{Punskaya+etal:2002} systems engineering \cite{Beck+Yuen:2004}\cite{Simon+Weber+Levrat:2007}\cite{Chulani+etal:1999}, and statistics in general \cite{Madigan+Raftery:1994}\cite{Hoeting+etal:1999}.

\section{Probability}
\label{sec:intro}
Logical statements can imply other logical statements.
Probability theory \cite{Berger:1985}\cite{Box&Tiao:1992}\cite{Jaynes:Book}\cite{MacKay:2003}\cite{Gregory:2005}\cite{Sivia&Skilling}\cite{Robert:2007}\cite{Bernardo+Smith:2009}\cite{Gelman+etal:2013}\cite{vonderLinden:2014} allows one to generalize the concept of implication by providing a measure of the degree of implication among logical statements \cite{Cox:1946}\cite{Cox:1961}.
More specifically, probability is a scalar measure that quantifies, within a topic of discourse, the degree to which one logical statement, representing a state of knowledge, implies another \cite{Knuth:measuring}\cite{Knuth&Skilling:2012}.\footnote{This is a relatively new interpretation of probability that has significant advantages over older concepts such as the \emph{frequency of occurrences of events}, the \emph{degree of truth} or the \emph{degree of belief}.} As a scalar measure, probability enables one to rank logical statements with respect to a given context or premise.

The utility of probability theory becomes apparent when one considers the degree to which a statement considering a set of several hypotheses or models, $M$, implies a joint statement proposing a particular model $m$ in conjunction with additional information or data, $d$, which we write as $P(m,d|M)$.  The product rule, which can be derived as a consequence of basic symmetries of Boolean logic \cite{Cox:1946}\cite{Cox:1961}\cite{Knuth:measuring}\cite{Knuth&Skilling:2012}, enable one to express this probability in two ways
\begin{eqnarray}
P(m,d|M)
&=& P(m|M) P(d|m,M) \\
&=& P(d|M) P(m|d,M).
\end{eqnarray}
These two expressions can be equated
\begin{equation} \label{eq:equated-product-rule}
P(m|M) P(d|m,M) = P(d|M) P(m|d,M)
\end{equation}
and rearranged resulting in the familiar Bayes' theorem
\begin{equation}
P(m|d,M) = P(m|M) \frac{P(d|m,M)}{P(d|M)},
\end{equation}
where the posterior probability $P(m|d,M)$ can be expressed in terms of the product of the prior probability $P(m|M)$ with a data-dependent term consisting of the ratio of the likelihood $P(d|m,M)$ to the evidence $P(d|M)$.  It is in this sense that one can think of Bayes' theorem as a learning rule where one's prior state of knowledge about the problem, represented by the prior probability, is updated by a data-dependent term resulting in a posterior probability that depends both on the prior state of knowledge as well as the data.

Both the prior probability and the likelihood must be assigned based on any and all additional information that one may possess about the problem. This is not a deficit or drawback of probability theory. Instead it is a strength since symmetries only serve to constrain manipulation of probabilities to the sum and product rules.  This leaves free the probability assignments resulting in a theory of inductive logic that can be applied to any particular inference problem. The dependence of these probabilities on problem-specific prior information is often indicated by including the symbol $I$ to the right of the solidus. \footnote{This notation goes back to Jaynes \cite{Jaynes:Book} and has been adopted in several prominent textbooks in the physical sciences \cite{Bretthorst:1988}\cite{Sivia&Skilling}\cite{Gregory:2005}\cite{vonderLinden:2014}.}
For example, this is done by writing the prior probability $P(m|M)$ as $P(m|M,I)$.

While the posterior probability over the space of models $M$ fully quantifies all that is known about the problem, it is often common practice to summarize what is known by focusing on a particular model $m$ that maximizes the posterior probability, such that this model is most implied by the data given the prior information.  Such a model is referred to as the most probable model or mode (within the context defined by the space of models $M$), or the maximum a posteriori (MAP) estimate.  Often the space of models $M$ to be considered is a parameterized space where each model $m$ is represented by a set of particular parameter values that act as coordinates in the space.  In this case, one can consider summarizing the posterior using the model given by the mean parameter values found using the posterior.  Either way, when the models in the space $M$ are parameterized, selecting a particular model given the data and prior information amounts to a parameter estimation problem.

The evidence, which in parameter estimation problems acts mainly as a normalization factor, can be obtained by summing or integrating (marginalizing) over all possible models $m$ in the set of models $M$
\begin{eqnarray}
P(d|M, I)
&=& \int{dm~P(m, d | M, I)}\\
\label{eq:evidence}&=& \int{dm~P(m | M, I) P(d | m, M, I)},
\end{eqnarray}
which is the reason that the evidence is often referred to as the marginal likelihood.

We can refer to a set of models, $M$, as a particular theory.  Given two competing theories $M_1$ or $M_2$ one can compare the posterior probability $P(M_1 | d, I)$ to the posterior probability $P(M_2 | d, I)$, where, among additional prior information, $I$ represents the fact that theories $M_1$ and $M_2$ are among those to be considered.  In general, both theories will result in non-zero probabilities.  However, the more probable theory can be determined by considering the ratio of their posterior probabilities.  We can examine this by considering the ratio of joint probabilities of the sets of models $M_1$ and $M_2$ and the data $d$ and then using the product rule to write the joint probability in two ways
\begin{eqnarray}
\frac{P(M_1, d | I)}{P(M_2, d | I)} &=& \frac{P(M_1, d | I)}{P(M_2, d | I)}\\
\frac{P(d | I) P(M_1|d, I)}{P(d | I) P(M_2|d, I)} &=& \frac{P(M_1 | I) P(d | M_1, I)}{P(M_2 | I) P(d | M_2, I)}\\
\frac{P(M_1|d, I)}{P(M_2|d, I)} &=& \frac{P(M_1 | I)}{P(M_2 | I)}\frac{P(d | M_1, I)}{P(d | M_2, I)}
\end{eqnarray}
so that the ratio of the posterior probabilities of the two theories is proportional to the ratio of their respective evidences.  The proportionality becomes an equality in the case where the prior probabilities of the two theories are equal.  This leads to the concept of the Bayes factor or odds ratio where we define
\begin{equation}
\mbox{OR} = \frac{P(d|M_1, I)}{P(d|M_2, I)}
\end{equation}
or, equivalently, the log odds ratio
\begin{equation} \label{eq:logOR}
\log{\mbox{OR}} = \log{P(d|M_1, I)} - \log{P(d|M_2, I)}.
\end{equation}
With this definition, we can write the ratio of posterior probabilities for the two different theories $M_1$ and $M_2$ in terms of the odds ratio
\begin{equation}
\frac{P(M_1|d, I)}{P(M_2|d, I)} = \frac{P(M_1 | I)}{P(M_2 | I)} \times \mbox{OR},
\end{equation}
where the two are equal when the ratio of the prior probabilities of the two theories are equal.

In the case of parameter estimation problems, the Bayesian evidence plays a relatively minor role as a normalization factor.  However, in problems where two theories are being tested against one another, which is often called a model selection problem\footnote{The terminology may be confusing since the term `model selection' seems to refer to the process of selecting a particular model; whereas, it refers to selecting one set of models, or theory, over another.}, the ratio of evidences is the relevant quantity to consider.  In some special cases, the integrals can be solved analytically as described in \cite{DeGroot:2004}\cite{Zellner+Siow:1980} and demonstrated below in Section \ref{sec:signal_detection}.

\section{Evidence, Model Order, and Priors}
It is instructive to consider how the evidence (\ref{eq:evidence}) varies as a function of the considered model order as well as the prior information one may possess about the model. We begin by considering a model consisting of a single parameter $x$, for which we have assigned a uniform prior probability over an interval $[x_{\min},x_{\max}]$ of width $\Delta x = x_{\max}-x_{\min}$.
  We can define the effective width $\delta{x} \le \Delta{x}$ of the likelihood over the prior range as
\begin{equation} \label{eq:effective_width}
\delta x \doteq \frac{1}{L_{\max}} \int^{x_{\max}}_{x_{\min}}dx~P(d|x,M,I),
\end{equation}
where $L_{\max}$ is the value of the likelihood $P(d|x,M,I)$ attained at the maximum likelihood estimate $x = \hat{x}$.
The evidence of the model amounts to
\begin{equation}
Z \equiv P(d|M, I) = \frac{1}{\Delta x} \int^{x_{\max}}_{x_{\min}}dx~P(d|x,M,I),
\end{equation}
which using the definition in (\ref{eq:effective_width}) can be conveniently expressed in terms of the prior width $\Delta x$ and effective likelihood width $\delta x$ by \cite[pp. 63-65]{Bretthorst:1988}
\begin{equation}
Z = L_{\max}\, \frac{\delta x}{\Delta x}.
\end{equation}
Thus we can write the evidence as a product of the maximum of the likelihood (the best achievable goodness-of-fit) and an Occam factor $W$:
\begin{equation}
Z = L_{\max}\, W
\end{equation}
where $0\le W\le 1$ is formally defined as
\begin{equation}
W = \frac{Z}{L_{\max}} = \int dx~P(x|M,I) \frac{P(d|x,M,I)}{L_{\max}}.
\end{equation}
For models with a single adjustable parameter the Occam factor is the ratio of the width of the likelihood over the prior range to the width of the prior: $W = {\delta x}/{\Delta x}$. For multiple model parameters this generalizes to the ratio of the volume occupied by those models that are compatible with both data and prior over the prior accessible volume.

%Maximum evidence is achieved if both widths are equal: our prior knowledge is fully compatible with the data, and the data do not tell us anything new about the model. In the general case, however, $\delta x \le \Delta x$.  %% Ridiculous!
By making the prior broader we pay in evidence. It is in this sense that Bayesian probability theory embodies Occam's razor: ``Entities are not to be multiplied without necessity.'' If we increase the flexibility of our model by the introduction of more model parameters, we reduce the Occam factor. Let's for simplicity assume that every additional parameter is also uniform over an interval of length $\Delta x$ and that there are $K$ such parameters $x_k$. Then beyond a certain model order $K$, we will achieve a perfect fit of the data upon which we cannot improve the likelihood any further. Because the Occam factor scales as $(\delta x/\Delta x)^K$, it will disfavor a further increase in model order.

Consider a Gaussian likelihood function, which is normalized so that it integrates to unity. If the data $d=\{d_1, \ldots, d_n\}$ are modeled as independent observations, the likelihood, assuming a standard deviation $\sigma$, is
\begin{equation}
P(d|x,M,I) = (2\pi\sigma^2)^{-n/2} \exp\left\{-\frac{n}{2\sigma^2} [(x-\overline{d})^2 + v] \right\}
\end{equation}
where $\overline{d} = \frac{1}{n} \sum_i d_i$ is the sample average and $v={\frac{1}{n} \sum_i (d_i - \overline{d})^2}$ the sample variance. Maximum likelihood is obtained at $\hat{x} = \overline{d}$ achieving a likelihood of $L_{\max} = ({e^{-v/\sigma^2}}/{2\pi\sigma^2})^{n/2}$.
The evidence is
\begin{equation}
P(d|M,I) = L_{\max}\sqrt{\frac{2\pi}{n}}\frac{\sigma}{\Delta x}\, \frac{\erf\bigl(\sqrt{\frac{n}{2}} \frac{x_{\max} -\overline{d}}{\sigma}\bigr) + \erf\bigl(\sqrt{\frac{n}{2}} \frac{\overline{d}-x_{\min}}{\sigma}\bigr)}{2}.
\end{equation}
For $\overline{d} \in [x_{\min}, x_{\max}]$ and $\sigma$ small or $n$ large, we can ignore the last factor involving the error function. The Occam factor is essentially $\sqrt{2\pi} \sigma/\Delta x\sqrt{n}$. If $\overline{d}$ falls outside the support of the prior ($\overline{d} < x_{\min}$ or $\overline{d} > d_{\max}$), the evidence decreases rapidly reflecting the discrepancy between our prior assumptions and the actual observations.

Let us compare a model $M_0$ that has no adjustable parameter and a model $M_1$ with a single adjustable parameter $x$ by computing the odds ratio:
\begin{equation}
\mbox{OR} = \frac{P(d|M_0, I)}{P(d|M_1, I)} \approx \frac{P(D| M_0, I)}{P(D|\hat{x}, M_1, I)}\, \frac{\Delta x}{\delta x}
\end{equation}
The odds ratio is comprised of two factors: the ratio of the likelihoods $$\frac{P(D| M_0, I)}{P(D|\hat{x}, M_1, I)}$$ and the Occam factor ${\Delta x}/{\delta x}$. The likelihood ratio is a classical statistic in frequentist model selection. If we only consider the likelihood ratio in model comparison problems, we fail to acknowledge the importance of Occam factors.

\section{Numerical Techniques}
In general, the evidence, which is found by integrating the prior times the likelihood (\ref{eq:evidence}) over the entire parameter space, cannot be solved analytically.\footnote{A rare exception is given by the first example presented in Section \ref{sec:signal_detection} where an analytical solution is obtained.}
This requires that we use numerical techniques to estimate the evidence. Straightforward estimation of the evidence integral directly from posterior sampling, such as in \cite{Chib:1995:gibbs}, proves to be quite challenging in general, especially in the case of multimodal distributions arising from mixture models or high-dimensional spaces.  While a number of sophisticated problem-specific techniques have been developed to handle such difficulties \cite{Berkhof+etal:2003}\cite{Trias+etal:2009}\cite{Chib+Srikanth:2010:tailored}, there is a need for more general widely-applicable techniques that require little to no fine tuning.

Other methods, such as \textit{Reversible Jump Markov Chain Monte Carlo} (RJMCMC) treat the model order as a model parameter \cite{Green:1995}\cite{Brooks+etal:2003:RJMCMC}\cite{Lopes+West:2004}.  However, such techniques typically encounter serious difficulties with inefficient model-switching moves.  The difficulties these more direct techniques experience are especially problematic in high-dimensional spaces and in problems where the likelihood calculations are expensive, such as in the case of large data sets or complex forward models.

This has resulted in the development of a rather sophisticated array of computational techniques.  Here we briefly review some of the more popular methods, pointing the interested readers to additional excellent resources and reviews, such as \cite{Kass+Raftery:1995} and \cite{Han+Carlin:2001}, and conclude with a focus on the more recent methods of nested sampling and its cousin MultiNest, which are used in three of the examples provided in the following section.

\subsection{Laplace Approximation}
The \textit{Laplace Approximation}, also known as the \textit{Saddle-Point Approximation} \cite{Daniels:1954saddlepoint}, is a simple and useful method for approximating a unimodal probability density function with a Gaussian \cite{MacKay:2003}\cite{Penny+Kiebel+Friston:2006:vb}\cite{vonToussaint:2011}\cite{vonderLinden:2014}. As such, the Laplace approximation forms the basis of more advanced techniques, such as Gull and MacKay's \textit{Evidence Framework} \cite{Gull:1989}\cite{Mackay:1992}.

Consider a function $p(x)$, which has a peak at $x = x_{0}$.  One can write the Taylor series expansion of the logarithm of the probability density $\ln p(x)$ about $x = x_{0}$ to second order as
\begin{equation}
\ln p(x) \simeq \ln p(x_{0}) + \left. \frac{d}{dx} \ln p(x) \right\vert_{x=x_{0}} (x-x_{0}) +  \left. \frac{1}{2}\frac{d^2}{dx^2} \ln p(x) \right\vert_{x=x_{0}} (x-x_{0})^2 + \ldots,
\end{equation}
which can be simplified to
\begin{equation} \label{eq:Laplace:lnP-tayor-series}
\ln p(x) \simeq \ln p(x_{0}) +  \left. \frac{1}{2}\frac{d^2}{dx^2} \ln p(x) \right\vert_{x=x_{0}} (x-x_{0})^2 + \ldots,
\end{equation}
since the first derivative of $\ln p(x)$ evaluated at the peak is zero
%: $\left. \frac{d}{dx} \ln p(x) \right\vert_{x=x_{0}} = 0$
.
By defining $\sigma^2$ to be minus the inverse of the local curvature at the peak
\begin{equation}
\sigma^2 = \left( \left. -\frac{1}{2}\frac{d^2}{dx^2} \ln p(x) \right\vert_{x=x_{0}} \right)^{-1},
\end{equation}
we can rewrite (\ref{eq:Laplace:lnP-tayor-series}) as
\begin{equation} \label{eq:Laplace:lnP-simplified}
\ln p(x) \simeq \ln p(x_{0}) -  \frac{1}{2\sigma^2}(x-x_{0})^2 + \ldots.
\end{equation}
Taking the exponential of both sides results in an un-normalized approximation for $p(x)$
\begin{equation}
p(x) \simeq p(x_{0}) \exp \left[ -\frac{1}{2\sigma^2}(x-x_{0})^2 \right],
\end{equation}
which would have as its normalization factor
\begin{equation} \label{eq:Laplace-one-dimensional-evidence}
Z = p(x_{0}) \sqrt{2 \pi \sigma^2}.
\end{equation}
If the function $p(x)$ is taken to be the product of the prior probability and the likelihood, then, the normalization factor (\ref{eq:Laplace-one-dimensional-evidence}) is an approximation of the evidence.

In $N$ dimensions, we expand the function $\ln p(\mathbf{x})$ as
\begin{equation}
\ln p(\mathbf{x}) \simeq \ln p(\mathbf{x}_{0}) -  \frac{1}{2}(\mathbf{x}-\mathbf{x}_{0})^{T} \mathbf{A} (\mathbf{x}-\mathbf{x}_{0}) + \ldots,
\end{equation}
where $\mathbf{A}$ is an $N \times N$ matrix, known as the Hessian, with matrix elements given by
\begin{equation}
A_{ij} = \left. - \frac{d^2}{dx_{i}dx_{j}} \ln p(\mathbf{x}) \right\vert_{x=x_{0}}.
\end{equation}
The approximation of $p(\mathbf{x})$ is then given by
\begin{equation}
p(\mathbf{x}) \simeq \frac{1}{Z} \exp{\left[ -\frac{1}{2} \mathbf{x}^T \mathbf{A} \mathbf{x} \right]}
\end{equation}
where the normalization factor is
\begin{equation}
Z = p(\mathbf{x}_{0}) \sqrt{\frac{(2 \pi)^N}{\det \mathbf{A}}}.
\end{equation}
Again, if the function $p(\mathbf{x})$ is defined by the product of the prior and the likelihood, then $Z$ is the approximation to the evidence.
This method requires that the peak of the distribution be identified and the Hessian estimated either analytically or numerically.

The Laplace approximation has been very useful in performing inference on latent Gaussian models, such as Gaussian processes \cite{Rasmussen+Williams:2006}.  The Integrated Nested Laplace Approximation (INLA) \cite{Rue+etal:2009}\cite{Martins+etal:2013} can be used to compute the posteriors of the model parameters in the case of structured additive regression models where the predictor depends on a sum of functions of a set of covariates, and the number of hyperparameters is small ($\leq 6$).  This is accomplished by setting up a grid of hyperparameter values where the posterior of the hyperparameter values given the data has been approximated using the Laplace approximation.  Then the Laplace approximation is used to compute the marginal posteriors given the data and the hyperparameter values across the grid.  The product of the hyperparameter posteriors (given the data) and the marginals (given the data and the hyperparameters) can then be numerically integrated over the hyperparameters to obtain the desired posterior marginals.  Another method to approximate the marginals based on expectation propagation \cite{Minka:2001} has been proposed by Cseke and Heskes \cite{Cseke+Heskes:2011}.  They demonstrated that this method is typically more accurate than INLA and works in cases where the Laplace approximation fails.

\subsection{Importance Sampling}
\textit{Importance Sampling} \cite{Neal:1993} allows one to find expectation values with respect to one distribution $p(x)$ by computing expectation values with respect to a second distribution $q(x)$
that is easier to sample from.  The expectation value of $f(x)$ with respect to $p(x)$ is given by
\begin{equation}
\left\langle f(x) \right\rangle_{p} = \frac{\int{f(x) p(x) \, dx}}{\int{p(x) \, dx}}.
\end{equation}
Note that one can write the distribution $p(x)$ as $\frac{p(x)}{q(x)} q(x)$ where the only theoretical requirement is that $q(x)$ must be non-zero wherever $p(x)$ is non-zero.  This allows one to rewrite the expectation value above as
\begin{align}
\left\langle f(x) \right\rangle_{p} &= \frac{\int{f(x) \frac{p(x)}{q(x)} q(x) \, dx}}{\int{\frac{p(x)}{q(x)} q(x) \, dx}} \\
&= \frac{ \left\langle f(x) \frac{p(x)}{q(x)} \right\rangle_{q} }{ \left\langle \frac{p(x)}{q(x)} \right\rangle_{q}},
\end{align}
which can be approximated with samples from $q(x)$ by
\begin{equation}
\left\langle f(x) \right\rangle_{p} \approx \frac{ \sum_{i=1}^{N}{f(x_i) \frac{p(x_i)}{q(x_i)}} }{ \sum_{i=1}^{N}{\frac{p(x_i)}{q(x_i)}}},
\end{equation}
where the samples $x = {x_1, x_2, \ldots, x_N}$ are drawn from $q(x)$.
This works well as long as the ratio defined by $p(x)/q(x)$ does not attain extreme values.
Importance sampling is a generally valid method useful even in cases where $q(x)$ is not Gaussian, as long as $q(x)$ is easier to sample from than $p(x)$ using techniques such as existing random number generators or MCMC.

Importance sampling can be used to compute ratios of evidence values in a similar fashion by writing \cite{Neal:1993}
\begin{equation}
\frac{Z_p}{Z_q} = \frac{\int{p(x) \, dx}}{\int{q(x) \, dx}}
\end{equation}
which can be written as
\begin{align}
\frac{Z_p}{Z_q} &= \frac{\int{\frac{p(x)}{q(x)} q(x) \, dx}}{\int{q(x) \, dx}} \\
&= \left\langle \frac{p(x)}{q(x)} \right\rangle_{q}
\end{align}
which can be approximated with samples from $q(x)$ by
\begin{equation}
\left\langle \frac{p(x)}{q(x)} \right\rangle_{q} \approx \frac{ \sum_{i=1}^{N}{\frac{p^2(x_i)}{q^2(x_i)}} }{ \sum_{i=1}^{N}{\frac{p(x_i)}{q(x_i)}}}.
\end{equation}
However, again the function $p(x)$ must be close to $q(x)$ to avoid extreme ratios, which will cause problems for the numeric integration.

\subsection{Analogy to Statistical Physics}
Techniques for evaluating the evidence can build on numerical methods in statistical physics because there is a close analogy between both fields. A key quantity in equilibrium statistical mechanics is the canonical partition function
\begin{equation}\label{eq:partitionfunction1}
Z(\beta) = \int dx~e^{-\beta E(x)}
\end{equation}
where $x$ are the configurational degrees of freedom of a system governed by the energy $E(x)$ and $\beta$ is the inverse temperature. Because $x$ is typically very high-dimensional, the partition function can only be evaluated numerically. Instead of computing $Z(\beta)$ directly by solving the high-dimensional integral (\ref{eq:partitionfunction1}), it is convenient to compute the \emph{Density of States} (DOS)
\begin{equation}
g(E) = \int dx~\delta(E - E(x))
\end{equation}
where $\delta$ is Dirac's delta function. The partition function and the DOS are linked via a Laplace transform
\begin{equation}\label{eq:partitionfunction2}
Z(\beta) = \int dE\, g(E) e^{-\beta E}.
\end{equation}
Therefore, knowing either of the two functions suffices to characterize equilibrium properties of the system and compute, for example, free energies and heat capacities.

In a Bayesian application, the model parameters $m$ play the role of the system's degrees of freedom and the negative log likelihood can be viewed as an energy function $E(m) = - \log P(d|m,M,I)$. For a given data set $d$, we write the DOS as
\begin{equation}
g(E) = \int dm~P(m|M,I) \delta[E - E(m)] %= \int dm~P(m|M,I) \delta[E + \log P(d|m,M,I)]
\end{equation}
The evidence can then be written as a one-dimensional integral over the DOS:
\begin{eqnarray}
P(d|M,I) &=& \int dE~g(E)\, e^{-E} \notag \\
&=& \int dm~P(m|M,I)\int dE~\delta[E + \log P(d|m, M, I)]\, e^{-E} \notag \\
&=& \int dm~P(m|M,I)\, P(d|m,M,I)
\end{eqnarray}
Therefore, knowledge of $g(E)$ allows us to compute the evidence in the same way as the canonical partition function (\ref{eq:partitionfunction2}) can be evaluated through a Laplace transform of the DOS \cite{Habeck:2012}.

Physics-inspired algorithms for evaluating the evidence aim to compute either the partition function $Z(\beta)$ at $\beta=1$ or the density of states. The previous class of methods comprises path sampling \cite{Gelman+Meng:1998}, parallel tempering \cite{Swendsen:1986,Geyer:1991}, annealed importance sampling \cite{Neal:2001} and other thermal methods that simulate a modified version of the posterior:
\begin{equation}
[P(d|m,M,I)]^\beta\, P(m|M,I)
\end{equation}
where the likelihood has been raised to a fractional power. By letting $\beta$ vary between zero and one, we can smoothly bridge between the prior and the posterior. A recent DOS-based algorithm called nested sampling \cite{Skilling:2004:nested}\cite{Skilling:2006:nested}\cite{Sivia&Skilling}\cite{vonderLinden:2014} is discussed in Section \ref{sec:nested}.

\subsection{Path Sampling and Thermodynamic Integration}
The method of \textit{path sampling} is based on the calculation of free energy differences in thermodynamics \cite{Gelman+Meng:1998}.  The method focuses on the estimation of the difference between the logarithm of two distributions $p_0$ and $p_1$, which depend on model parameters.  One can connect the two distributions by a ``path'' through a space of distributions by defining what is called the geometric path
\begin{equation}\label{eq:geometricbridge}
p(x|\beta) \propto p_0(x)^{1-\beta} p_1(x)^{\beta}
\end{equation}
where the parameter $\beta$ can vary freely from $\beta=0$ to $\beta=1$ so that at the endpoints we have that $p(x|\beta=0) = p_0(x)$ and $p(x|\beta=1) = p_1(x)$. By letting $E = \log [p_0/p_1]$ we can establish a direct relation with the canonical ensemble; the normalizing constant is the partition function:
\begin{align}
Z(\beta) &= \int dx~p_0(x)^{1-\beta} p_1(x)^{\beta} \notag \\
&= \int dx~p_0(x)\, e^{-\beta E(x)}.
\end{align}
The log partition function can be estimated using samples from $p(x|\beta)$ in the following way. We have
\begin{align}
\partial_\beta \log Z(\beta) &= - \frac{1}{Z(\beta)}  \int dx~E(x)\, p_0(x)\, e^{-\beta E(x)} \notag \\
&= \left\langle \log [p_1/p_0] \right\rangle_\beta
\end{align}
where $\left\langle \cdot \right\rangle_\beta$ denotes the expectation with respect to the bridging distribution $p(x|\beta)$. Integration of the previous equation yields
\begin{align}
\log[ Z(1)/Z(0)] &= \int_0^1 d\beta~\partial_\beta \log Z(\beta) \notag \\
&= \int_0^1 d\beta~\left\langle \log [p_1/p_0] \right\rangle_\beta.
\end{align}
By choosing a finely spaced $\beta$-path we can approximate the ratio of the normalization constants $Z(1)/Z(0)$ by a sum over the expected energy $\log[ p_0/p_1]$ (log likelihood ratio) over each of the bridging distributions:
\begin{equation}\label{eq:thermodynamicintegration}
\log[ Z(1)/Z(0)] \approx \sum_i \left\langle \log [p_1/p_0] \right\rangle_{\beta_i} (\beta_{i+1} - \beta_i).
\end{equation}
This approach is called \textit{thermodynamic integration}. It is also possible to estimate the DOS from samples produced along a thermal path bridging between the prior and posterior and thereby obtain an alternative estimate of the evidence that is sometimes more accurate than thermodynamic integration \cite{Habeck:2012,Habeck:2012b}.

If we choose $p_0(m) = P(m|M,I)$ and $p_1(m) = P(m|M,I)\, P(d|m,M,I)$, we can use path sampling in combination with thermodynamic integration to obtain the log-evidence because $Z(0) = \int dm~p_0(m) = 1$ and $Z(1) = \int dm~p_1(m) = P(d|M,I)$.  In case we want to compare two models $M_1, M_2$ that share the same parameters $m$, we can use thermodynamic integration to estimate the log odds ratio (\ref{eq:logOR}) by defining $p_{i-1}(m) = P(m|M_i,I)\, P(d|m,M_i,I)$ ($i=1,2$) and sampling from the following family of bridging distributions
\begin{equation}
p(x|\beta) \propto [P(m|M_1,I)\, P(d|m,M_1,I)]^{1-\beta}\, [P(m|M_2,I)\, P(d|m,M_2,I)]^{\beta}
\end{equation}
For the special case that both models also share the same prior, $P(m|M_1,I) = P(m|M_2,I) = P(m|I)$, this simplifies to
\begin{equation}
p(m|\beta) \propto P(m|I)\, [P(d|m,M_1,I)]^{1-\beta}\, [P(d|m,M_2,I)]^{\beta} .
\end{equation}
By drawing models from the mixed posterior $p(m|\beta)$ the log odds ratio can be computed directly using thermodynamic integration. An open problem relevant to all thermal methods using a geometric path (\ref{eq:geometricbridge}) is where to place the intermediate distributions. This becomes increasingly difficult for complex systems that show a phase transition.

\textit{Ensemble Annealing} \cite{Habeck:2015}, a variant of \textit{simulated annealing} \cite{Kirkpatrick+etal:1983}, aims to circumvent this problem by constructing an optimal temperature schedule in the course of the simulation. This is achieved by controlling the relative entropy between successive intermediate distributions: After simulating the system at a current temperature, the new temperature is chosen such that the estimated relative entropy between the current and the new distribution is constant. Ensemble annealing can be viewed as a generalization of nested sampling (see Section \ref{sec:nested}) to general families of bridging distributions such as the geometric path (\ref{eq:geometricbridge}). Ensemble annealing has been applied to various systems showing first- and second-order phase transitions such as Ising, Potts, and protein models \cite{Habeck:2015}.

\subsection{Annealed Importance Sampling}
\textit{Annealed Importance Sampling} (AIS) \cite{Neal:2001} is closely related to other annealing methods such as simulated annealing but does not rely on thermodynamic integration. AIS generates multiple independent sequences of states $\{ x_0^{(j)},x_1^{(j)},\ldots,x_{i}^{(j)},\ldots\}$ where ${x}_i^{(j)}$ is a sample from the $i$-th intermediate distribution $p_i$ bridging between the initial distribution $p_0$ and the destination distribution $p_1$. For example, in case we are using the geometric bridge (\ref{eq:geometricbridge}) the states follow
\begin{equation}
x_i^{(j)} \sim p_0(x)^{1-\beta_i} p_1(x)^{\beta_i}
\end{equation}
where the superscript $j$ enumerates the sequences. Sampling of $x_i^{(j)}$ is typically achieved by starting a Markov chain sampler from the precursor state $x_{i-1}^{(j)}$. Each of the generated sequences is assigned an importance weight
\begin{equation}\label{eq:aisweights}
w^{(j)} = \prod_{i} \frac{p_{i+1}({x}_{i}^{(j)}) }{ p_{i}({x}_{i}^{(j)}) }\,.
\end{equation}
Neal has shown \cite{Neal:2001} that the average of the importance weights is an unbiased estimator of the ratio of the normalizing constants:
\begin{equation}\label{eq:ais}
  \frac{Z(1)}{Z(0)} \approx \frac{1}{M} \sum_{j=1}^M w^{(j)} = \frac{1}{M} \sum_{j=1}^M \prod_i \frac{p_{i+1}({x}_{i}^{(j)}) }{ p_{i}({x}_{i}^{(j)}) }\,.
\end{equation}
It is important to note that the annealing sequence is simulated multiple times, and that the partition function is obtained from the importance weights $w^{(j)}$ by an {arithmetic} average rather than a geometric average. For the special case of the geometric bridge, the AIS estimator is
\begin{equation}
  \frac{Z(1)}{Z(0)} \approx \frac{1}{M} \sum_j \exp\biggl\{\sum_i (\beta_{i+1}-\beta_i)\, \log[p_1({x}_{i}^{(j)})/p_0({x}_{i}^{(j)})] \biggr\}\,.
\end{equation}

On the other hand, if we apply thermodynamic integration [Eq. (\ref{eq:thermodynamicintegration})] to the sequences sampled during AIS, we obtain
\begin{equation}
\log [{Z(1)}/{Z(0)}] \approx \sum_i (\beta_{i+1} - \beta_i)\, \frac{1}{M} \sum_j \log[p_1({x}_{i}^{(j)})/p_0({x}_{i}^{(j)})]\, .
\end{equation}
Both estimators are closely related but not identical. To see this, let us rewrite the estimate obtained by thermodynamic integration:
\begin{eqnarray}
\frac{Z(1)}{Z(0)} &\approx& \exp\biggl\{ \frac{1}{M} \sum_j \sum_i (\beta_{i+1} - \beta_i)\, \log[p_1({x}_{i}^{(j)})/p_0({x}_{i}^{(j)})] \biggr\} \\
&\approx& \exp\biggl\{ \frac{1}{M} \sum_j \log w^{(j)} \biggr\} = \biggl( \prod_j w^{(j)} \biggr)^{1/M}\, .
\end{eqnarray}
This shows that AIS estimates the ratio of partition functions by an arithmetic average over the importance weights, whereas thermodynamic integration averages the importance weights $w^{(j)}$ geometrically. Neal's analysis as well as results from non-equilibrium thermodynamics (e.g. \cite{Crooks:1999}) show that the AIS estimator is valid even if the sequences of states are not in equilibrium.

\subsection{Variational Bayes} \label{sec:vb}
Another technique called \textit{Ensemble Learning} \cite{Hinton+VanCamp:1993}\cite{Mackay:1995:ensemble}\cite{Valpola+Karhunen:2002}, or \textit{Variational Bayes} (VB) \cite{Waterhouse+etal:1996}\cite{Attias:1999:vb}\cite{Lawrence+Bishop:2000:vb}\cite{Sato:2001:vb}\cite{Roberts+Penny:2002:vb}\cite{Penny+Kiebel+Friston:2006:vb}\cite{Smidl+Quinn:2006:vb}, is named after Feynman's variational free energy method in statistical mechanics \cite{Feynman:1972}.  As such, it is yet another example of how methods developed in thermodynamics and statistical mechanics have had an impact in machine learning and inference.

We consider a normalized probability density $Q(m)$ on the set of model parameters $m$, such that
\begin{equation}
\int{dm \, Q(m)} = 1.
\end{equation}
While not obviously useful, the log-evidence can be written as
\begin{equation}
\log{P(M|I)} = \int{dm \, Q(m) \log{P(M|I)}}.
\end{equation}
Using the product rule, this can be written as
\begin{align}
\log{P(M|I)} &= \int{dm \, Q(m) \log{\frac{P(M,m|I)}{P(m|M,I)}}}\\
&= \int{dm \, Q(m) \log\left[{\frac{P(M,m|I) Q(m)}{P(m|M,I) Q(m)}}\right]}.
\end{align}
This expression can be broken up into the sum of the negative free energy
\begin{equation} \label{eq:NegFreeEnergy}
F(Q(m),P(M,m|I)) = \int{dm \, Q(m) \log{\frac{P(M,m|I)}{Q(m)}} }
\end{equation}
and the Kullback-Leibler (KL) divergence
\begin{equation}
KL[ Q(m) \| P(m|M,I) ]=  \int{dm \, Q(m) \log{\frac{Q(m)}{P(m|M,I)}} }
\end{equation}
by
\begin{equation} \label{eq:VariationalBayes}
\log{P(M|I)} = F(Q(m),P(M,m|I)) + KL[ Q(m) \| P(m|M,I) ],
\end{equation}
which is the critical concept behind variational Bayes.

The properties of the KL divergence expose an important relationship between the negative free energy and the evidence.  First, the KL divergence is zero when the density $Q(m)$ is equal to the posterior $Q(m) = P(m|M,I)$.  For this reason, $Q(m)$ is referred to as the approximate posterior. Furthermore, since the KL divergence is always non-negative, we have that
\begin{equation}
\log{P(M|I)} = \max_{Q}F(Q(m),P(M,m|I)) \geq F(Q(m),P(M,m|I))
\end{equation}
so that the negative free energy is a lower bound to the log-evidence.

The main idea is to vary the density $Q(m)$ (approximate posterior) so that it approaches the posterior $P(m|M,I)$.  One cannot do this directly through the KL divergence since the evidence, which is the normalization factor for the posterior, is not known.  Instead, by maximizing the negative free energy in (\ref{eq:VariationalBayes}), which is the same as minimizing the free energy, the negative free energy approaches the log-evidence and the approximate posterior $Q(m)$ approaches the posterior.  However, this presents a technical difficulty in that the integral for the negative free energy (\ref{eq:NegFreeEnergy}) will not be analytically solvable for arbitrary $Q(m)$.  The approach generally taken involves a concept from the mean field approximation in statistical mechanics \cite{McComb:2004:renormalization} where a non-factorizable function is replaced by one that is factorizable
\begin{equation}
Q(m) = Q(m_{0})Q(m_{1})
\end{equation}
where the set of model parameters $m$ can be divided into two disjoint sets $m_{0}$ and $m_{1}$ so that $m_{0} \cap m_{1} = \varnothing$ and $m_{0} \cup m_{1} = m$.

The negative free energy (\ref{eq:NegFreeEnergy}) can then be written as \cite{Penny+Kiebel+Friston:2006:vb}
\begin{align}
F &= \int{dm \, Q(m) \log{\frac{P(M,m|I)}{Q(m)}} } \notag \\
&= \int\int{dm_{0} \, dm_{1} \, Q(m_{0})Q(m_{1}) \log{\frac{P(M,m_{0},m_{1}|I)}{Q(m_{0})Q(m_{1})}} } \notag \\
&= \int{dm_{0} \,  Q(m_{0}) \left[ \int{dm_{1} \, Q(m_{1}) \log{P(M,m_{0},m_{1}|I)} }\right] } \notag \\
& \qquad \qquad \qquad \qquad \qquad \qquad  \qquad - \int{dm_{0} \,  Q(m_{0}) \log{Q(m_{0})} } + C  \notag \\
&= \int{dm_{0} \,  Q(m_{0}) I(m_{0}) } - \int{dm_{0} \,   Q(m_{0}) \log{Q(m_{0})} } + C  \notag
\end{align}
where the constant $C$ consists of terms that do not depend on $Q(m_{0})$ and
\begin{equation}
I(m_{0}) = \int{dm_{1} \, Q(m_{1}) \log{P(M,m_{0},m_{1}|I)} }.
\end{equation}

The negative free energy can then be expressed in terms of a KL-divergence by writing $I(m_{0}) = \log(\exp(I(m_{0})))$
\begin{equation}\label{eq:vb-mean-field-negative-free-energy}
F = KL[ Q(m_{0}) \| \exp(I(m_{0})) ] + C,
\end{equation}
which is minimized when
\begin{equation}
Q(m_{0}) \propto \exp(I(m_{0})).
\end{equation}
This implies that not only can the posterior be approximated with $Q(m)$, but also the analytic form of the component posteriors can be determined.  This is known as the \textit{free-form approximation} \cite{Penny+Kiebel+Friston:2006:vb}, which applies, in general, to the conjugate exponential family of distributions \cite{Attias:1999:vb}\cite{Ghahramani+Beal:2001:vb}\cite{Winn+Bishop:2005:vb}, and can be extended to non-conjugate distributions \cite{Attias:1999:vb}\cite{Jaakkola+Jordan:1996}.

Since the negative free energy (\ref{eq:vb-mean-field-negative-free-energy}) is a lower bound to the log-evidence, the log-evidence can be estimated by minimizing the negative free energy, so that the approximate posterior $Q(m)$ approaches the posterior.

\subsection{Nested Sampling} \label{sec:nested}
\begin{figure}
\centering
\includegraphics[scale = 0.35]{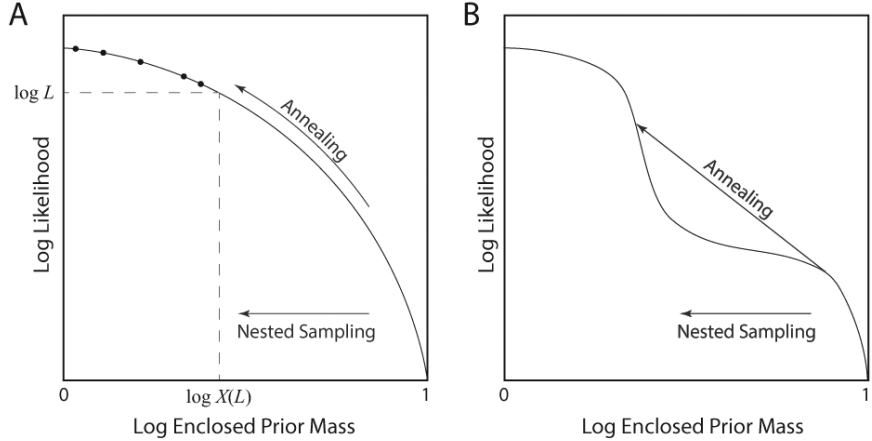}
\caption{A. An illustration of the amount of log prior mass $\log{X(L)}$ with a log likelihood greater than $\log{L}$.  Nested sampling relies on forming a nested set of likelihood boundaries within which the $N$ walkers are uniformly distributed. As such, nested sampling contracts the prior volume to higher likelihood at a steady rate based on $\log X$ \cite{Sivia&Skilling}.  On the other hand, tempering methods, such as simulated annealing, advance on high likelihood regions by following the slope $\frac{d\log{L}}{d\log{X}}$ of the curve.   B. Tempering methods, which slowly turn on the likelihood $L^{\beta}$ with the inverse temperature parameter $\beta$, are designed to follow the concave hull of the log likelihood. In situations where the slope becomes convex, one must jump from one phase (local maximum in evidence mass) to another, which is why tempering methods typically fail at phase transitions.  Nested sampling, which contracts the prior volume, is not hampered by such features in the log likelihood curve as a function of prior mass.}
\label{fig:Nested-Sampling}
\end{figure}

Nested sampling \cite{Skilling:2004:nested}\cite{Skilling:2006:nested}\cite{Sivia&Skilling}\cite{vonderLinden:2014} relies on stochastic integration to numerically compute the evidence of the posterior probability. In contrast to the thermal algorithms discussed so far, nested sampling aims to estimate the DOS or rather its cumulative distribution function
\begin{align}
X(L) &= \int^{-\log L}_{-\infty} dE~g(E) \notag \\
&= \int_{P(d|m,M,I) > L} dm\, P(m|M,I) \label{eq:priormass}
\end{align}
which calculates the prior mass $X\in[0,1]$ contained in the likelihood contour $P(d|m,M,I) > L\equiv e^{-E}$. We can now write the evidence integral as
\begin{align}
Z &= \int_{-\infty}^\infty dE~g(E) e^{-E} \notag \\
&= \int_0^1 dX~L(X) \notag \\
&\approx \sum_i L_i (X_{i-1} - X_i) \label{eq:evidenceNS}
\end{align}
where the likelihood $L(X)$ is understood as a function of the cumulative DOS or prior mass (\ref{eq:priormass}). Because $L(X)$ is unknown for general inference problems, we have to estimate it. Nested sampling does this by estimating its inverse function $X(L)$ using $N$ walkers that explore the prior constrained by a lower/upper bound on the likelihood/energy (Figure \ref{fig:Nested-Sampling}A). Since $X$ decreases monotonically in likelihood, we can sort the unknown prior masses associated with each walker by sorting them according to likelihood. The walker with worst likelihood will enclose the largest prior mass. The maximum mass can be estimated using order statistics:
\begin{equation}\label{eq:orderstatistics}
X_{\max} \sim N\, \frac{X_{\max}^{N-1}}{X(L)}
\end{equation}
where the walkers have been numbered such that they increase in likelihood $L_1<L_2<\ldots<L_N$ and thus $X_{\max}\equiv X_1> X_2 > \ldots > X_N$. The worst likelihood $L_1$ will define the lower likelihood bound in the next iteration. Walkers $2$ to $N$ will, by construction, already attain states that are also valid samples from the prior truncated at $L_1$ such that we only have to replace the first walker. This can be done by randomly selecting one among the $N-1$ surviving states and evolving it within the new contour $L_1$ using a Monte Carlo procedure. The initial states are obtained by sampling from the prior (i.e. the lower bound on the likelihood is zero); the associated mass is $X(0) = 1$.

Rather than increasingly giving the data more and more weight as is done in thermal approaches, nested sampling focuses on states with high posterior probability by constructing a sequence of nested priors restricted to higher and higher likelihood regions. Thereby nested sampling locates the relevant states that contribute strongly to the evidence integral and simultaneously constructs an optimal sequence of likelihood contours. In path sampling, the geometric path must be typically chosen by the user; nested sampling elegantly circumvents this problem.

Another advantage of nested sampling is that each likelihood bound in the nested sequence compresses the prior volume by approximately the same factor, which allows nested sampling to handle first order phase transitions.  In contrast, tempering methods such as simulated annealing and parallel tempering compress based on steps in temperature ($\propto L^{1/T}$), and as a result they typically fail at phase transitions (Figure \ref{fig:Nested-Sampling}B).

Often a practical difficulty of applying nested sampling is the requirement to sample from the prior subject to a hard constraint on the likelihood.  Mukherjee et al. \cite{Mukherjee+etal:2006} developed a version of nested sampling that fits an enlarged ellipse around the walkers and samples uniformly from that ellipse until a sample is drawn that has a likelihood exceeding the old minimum.  Sampling within the hard constraint is also made difficult when the distribution is multi-modal.  MultiNest \cite{Feroz+Hobson:2008}\cite{Feroz+etal:2009}\cite{Feroz:2013} was developed to handle multi-modal distributions by using K-means clustering to cluster the walkers into a set of ellipsoids.  At each iteration, MultiNest replaces the walker with the worst likelihood by a new walker generated by randomly selecting an ellipsoid (uniformly) and sampling uniformly from within the bounds of that ellipsoid.  These ellipsoids serve to allow one to detect and characterize multiple peaks in the distribution. However, the method has two drawbacks in which accurate K-means clustering limits the dimensionality of the problem to tens of parameters, and the elliptical regions may not always cover the high likelihood regions of the parameter space.

Other variants of nested sampling couple the technique with Hamilton Monte Carlo \cite{Betancourt:2010} or Galilean Monte Carlo \cite{Skilling:2012:GMC}\cite{Feroz+Skilling:2013}\cite{Goggans+Henderson+Xiang:2013}, which sample within the hard likelihood constraint by considering the step size to be determined by some particle dynamics depending on the particle velocity, and using that velocity and likelihood gradient to reflect off of the hard likelihood boundary.  This has been demonstrated to result in improved exploration in cases of multi-modal distributions and distributions with curved degeneracies.

Another possible way to facilitate sampling from within the hard likelihood constraint is to introduce additional ``demon'' variables that smooth the constraint boundary and push the walkers away from it \cite{Habeck:2014}. This approach can help to solve complex inference problems as they arise, for example, in protein structure determination, at the expense of introducing additional algorithmic parameters.

Diffusive Nested Sampling \cite{Brewer+etal:2011} is a variant of nested sampling that monitors the log likelihood values during the MCMC steps and creates nested levels such that each level covers approximately $e^{-1}$ of the prior mass of the previous level.  This allows the relative enclosed prior mass of the nested levels to be estimated more accurately than in nested sampling.  Samples are then obtained from a weighted mixture of the current level and the previous levels so that a mixture of levels is diffusively explored facilitating travel between isolated modes and allowing a more refined estimate of the log evidence.

\section{Practical Examples}
In this section we consider a set of four practical examples where the Bayesian evidence is both calculated and used in different ways.    The purpose of this section is not to compare one computational method against another, since given the large number of techniques available, this would require a more extensive treatment.  Instead, the goal is to demonstrate the utility of Bayesian model selection in several examples both relevant to signal processing and spanning the domain sciences.

The first example focuses on the problem of signal detection where the evidence, which is computed analytically, is used to test between two models: signal present and signal absent.  The second example focuses on using the evidence, estimated numerically by nested sampling, to select the model order of a Gaussian mixture model of the spatial sensitivity function of a light sensor.  The third example relies on the application of the evidence, estimated using MultiNest, to select among a set of exoplanet models each exhibiting different combinations of photometric effects. The final example selects a molecular mechanics force field approximately describing atomic interactions in proteins by computing the evidence of nuclear magnetic resonance (NMR) data.

\subsection{Signal Detection}
\label{sec:signal_detection}
In this example, based on the work by Mubeen and Knuth \cite{Mubeen+Knuth:2014}, we consider a practical signal detection problem where the log odds-ratio can be analytically derived.  The result is a novel signal detection filter that outperforms correlation-based detection methods in the case where both the noise variance and the variance in the overall signal amplitude is known.  While this detection filter was originally designed to be used in brain-computer interface (BCI) applications, it is applicable to signal detection in general (with slight modification).

We consider the problem of detecting a stereotypic signal, $s(t)$, which is modeled by a time-series with $T$ time points.  This signal has the potential to be recorded from $M$ detector channels with various (potentially negative) coupling weights $C_m$ where the index $m$ refers to the $m^{th}$ channel.  Last, and perhaps more specific to the BCI problem, we consider that the overall amplitude of the emitted signal waveshape $s(t)$ can vary.  This is modeled using a positive-valued amplitude parameter $\alpha$, which is the only free parameter as it is assumed that the coupling weights $C_m$ and the signal waveshape $s(t)$ are known.

There are two states to be considered: \emph{signal absent} (null hypothesis) and \emph{signal present}.  We model the signal absent state as noise only
\begin{equation}\label{eq:bmi:signal-absent}
M_{N} \quad : \quad x_{m}(t) = n_{m}(t)
\end{equation}
where $M_N$ denotes the ``noise-only'' model, $x_{m}(t)$ denotes the signal time-series recorded in the $m^{th}$ channel and $n_m(t)$ refers to the noise signal associated with the $m^{th}$ channel.  The signal present state is modeled as signal plus noise by
\begin{equation}\label{eq:bmi:signal-present}
M_{S+N} \quad : \quad x_{m}(t) = \alpha C_{m} s(t) + n_{m}(t)
\end{equation}
where the symbol $M_{S+N}$ denotes the ``signal-plus-noise'' model and $\alpha$ is the amplitude of the signal $s(t)$, which is coupled to each of the $m$ detectors with weights $C_m$.

The \emph{odds-ratio} can be written as the ratio of evidences
\begin{equation}
\mbox{OR} = \frac{P(X | M_{S+N}, I)}{P(X | M_{N}, I)} \equiv \frac{Z_{S+N}}{Z_{N}} \label{eq:OR}
\end{equation}
where $X$ represents the available data, which here will be the recorded time series vector $\mathbf{x}(t) = \{x_1(t), x_2(t), \ldots, x_M(t) \}$, and $I$ represents any relevant prior information including the coupling weights $C_m$ and the signal waveshape $s(t)$.  The two evidence values can be written as
\begin{align}
Z_{N} &= P(X | M_{N}, I)  \\
&= P(\mathbf{x}(t) | \mathbf{n}(t), I) \label{eq:likelihoodZn}
\end{align}
and
\begin{align}
Z_{S+N} &= P(X | M_{S+N}, I)  \\
&= \int^{\alpha_{\max}}_{\alpha_{\min}}{d\alpha \, P(\alpha | I) P(\mathbf{x}(t) | \mathbf{n}(t), I)} \label{eq:Z_S+N}
\end{align}
where the latter is marginalized over the amplitude range $[\alpha_{\min}, \alpha_{\max}]$ of the signal $\alpha$ since we only care to detect the signal. Here $P(\alpha | I)$ represents the prior probability for the amplitude parameter $\alpha$.  Note also that $\mathbf{x}(t)$ and $\mathbf{n}(t)$ without subscripts refer to the vector of time series over each of the detector channels.

Assuming that the noise signals $\mathbf{n}(t)$ have identical characteristics in each channel, we assign a Gaussian likelihood with a standard deviation of $\sigma_n$ to both models.  Note that this is not quite the same as assuming that the signals are Gaussian distributed, but rather this is the maximum entropy assignment where both the mean and squared deviation from the mean are known to be relevant quantities.  For the ``noise-only'' model there are no model parameters and the likelihood is equal to the evidence (\ref{eq:likelihoodZn})
\begin{equation}
Z_{N} = (2 \pi {\sigma_n}^2)^{-MT/2} \exp \left[ {-\frac{1}{2 {\sigma_{n}}^2} \sum_{m=1}^{M}{ \sum_{t=1}^{T}{ {x_{m}}^2(t) } } } \right].
\end{equation}
In the ``signal-plus-noise'' model we have the Gaussian likelihood
\begin{multline}\label{eq:likelihood:S+N}
P(\mathbf{x}(t) | \alpha, n(t), I) = \\
(2 \pi {\sigma_n}^2)^{-MT/2} \exp \left[ {-\frac{1}{2 {\sigma_{n}}^2} \sum_{m=1}^{M}{ \sum_{t=1}^{T}{ (x_{m}(t) - \alpha C_{m} s(t))^2 } } } \right].
\end{multline}
By assigning a (potentially-truncated) Gaussian prior to the amplitude parameter $\alpha$,
%with mean $\hat{\alpha}$ and standard deviation $\sigma_\alpha$ we have
\begin{equation}\label{eq:BMI:truncated-prior-alpha}
P(\alpha | I) = \frac{1}{Z_{\alpha}} \exp \left[ {-\frac{1}{2 {\sigma_{\alpha}}^2} ({\alpha - \hat{\alpha}})^2 } \right],
\end{equation}
with a normalization constant $Z_{\alpha}$ given by
\begin{equation}
Z_{\alpha} = \int_{\alpha_{\min}}^{\alpha_{\max}}d\alpha~\exp\left[-\frac{1}{2\sigma_\alpha^2} (\alpha - \hat{\alpha})^2\right],  \label{eq:Za_int}
\end{equation}
one can integrate the likelihood (\ref{eq:likelihood:S+N}) to find the evidence of the ``signal-plus-noise'' model (\ref{eq:Z_S+N}).
%\begin{equation}\label{eq:BMI:truncated-prior-alpha}   % OLD WITH MISTAKE IN NORMALIZATION
%P(\alpha | I) = (2 \pi {\sigma_\alpha}^2)^{-1/2} \exp \left[ {-\frac{1}{2 {\sigma_{\alpha}}^2} ({\alpha - \hat{\alpha}})^2 } \right],
%\end{equation}
%which along with (\ref{eq:likelihood:S+N}) will allow us to perform the integration in (\ref{eq:Z_S+N}).

By defining
\begin{equation}
D = S^2 + \sum_{m=1}^{M}{ \sum_{t=1}^{T}{ {C_{m}}^2 s^2(t) } }
\end{equation}
\begin{equation}
E = S^2 \hat{\alpha} + \sum_{m=1}^{M}{ \sum_{t=1}^{T}{ C_{m} x_{m}(t) s(t) } } \label{eq:E}
\end{equation}
\begin{equation} \label{eq:F}
F = S^2 \hat{\alpha}^2 + \sum_{m=1}^{M}{ \sum_{t=1}^{T}{ {x_{m}}^2(t) } },
\end{equation}
where
\begin{equation}
S^2 = \frac{{\sigma_n}^2}{{\sigma_{\alpha}}^2},
\end{equation}
we can   complete the square in the exponent and write the odds ratio as
\begin{align}
\frac{Z_{S+N}}{Z_{N}} &= \frac{\int^{\alpha_{\max}}_{\alpha_{\min}}{d\alpha \, P(\alpha | I) P(\mathbf{x}(t) | \mathbf{n}(t), I)}}{Z_{N}} \\
&= {\exp \left[ {-\frac{1}{2\sigma_n^2} (S^2\hat{\alpha}^2 - E^2/D)} \right] } \frac{Z_d}{Z_\alpha} \label{eq:bmi:odds-ratio-initial}
\end{align}
where
\begin{equation}
Z_d = \int_{\alpha_{\min}}^{\alpha_{\max}}d\alpha~\exp\left[-\frac{D}{2\sigma_n^2} (\alpha - E/D)^2\right] \label{eq:Zd_int}.
\end{equation}

In general, these Gaussian integrals result in solutions involving the error function ($\erf$) \cite{Abramowitz&Stegun:1972}
\begin{equation}
\int_a^b dx~e^{-\frac{1}{2\sigma^2} (x-\mu)^2} = \frac{\sqrt{2\pi\sigma^2}}{2} \left[\mathrm{erf}\biggl(\frac{b-\mu}{\sqrt{2}\,\sigma}\biggr) + \mathrm{erf}\biggl(\frac{\mu-a}{\sqrt{2}\,\sigma}\biggr) \right].
\end{equation}
If we restrict the signal amplitude to being positive, we have that $\alpha_{\min} = 0$ and $\alpha_{\max} = +\infty$
and the integrals (\ref{eq:Zd_int}) and (\ref{eq:Za_int}) become
\begin{equation}
Z_d = \frac{\sqrt{2\pi\sigma_n^2/D}}{2} \left[1 + \mathrm{erf}\biggl(\frac{E}{\sqrt{2D}\,\sigma_n}\biggr) \right]
\end{equation}
and
\begin{equation}
Z_\alpha = \frac{\sqrt{2\pi\sigma_\alpha^2}}{2} \left[1 + \mathrm{erf}\biggl(\frac{\hat{\alpha}}{\sqrt{2}\,\sigma_\alpha}\biggr) \right]
\end{equation}
resulting in the log odds ratio
\begin{multline}
\log \mbox{OR}_{+} = \\
\frac{1}{2} \left[\left(\frac{E^2}{D\sigma_n^2} - \frac{\hat{\alpha}^2}{\sigma_\alpha^2}\right) + \log\left(\frac{ S^2}{D}\right)\right] + \log\left(\frac{1 + \mathrm{erf}\biggl(\frac{E}{\sqrt{2D}\,\sigma_n}\biggr)}{1 + \mathrm{erf}\biggl(\frac{\hat{\alpha}}{\sqrt{2}\,\sigma_\alpha}\biggr)}\right),
\end{multline}
where the subscript $+$ indicates that the signal amplitude $\alpha$ is assumed to be positive.

However, if we consider allowing $\alpha$ to vary over the entire real line by setting $\alpha_{\min} = -\infty$ and $\alpha_{\max} = +\infty$ we find that
\begin{equation}
Z_d = \sqrt{2\pi\sigma_n^2/D}
\end{equation}
and
\begin{equation}
Z_\alpha = \sqrt{2\pi\sigma_\alpha^2},
\end{equation}
which gives a simpler log odds ratio which lacks the term with the $\erf$ functions
\begin{equation}
\log \mbox{OR}_{\pm} = \frac{1}{2} \left[\left(\frac{E^2}{D\sigma_n^2} - \frac{\hat{\alpha}^2}{\sigma_\alpha^2}\right) + \log\left(\frac{ S^2}{D}\right)\right],
\end{equation}
where the subscript $\pm$ indicates that the signal amplitude $\alpha$ ranges from $-\infty$ to $\infty$.

The expression $E$ (\ref{eq:E}) contains the cross-correlation term, which is what is typically used for the detection of a target signal in ongoing recordings.  The log OR detection filters incorporate more information that leads to extra terms, which serve to aid in target signal detection.

\begin{figure}
\centering
\includegraphics[scale = 0.39]{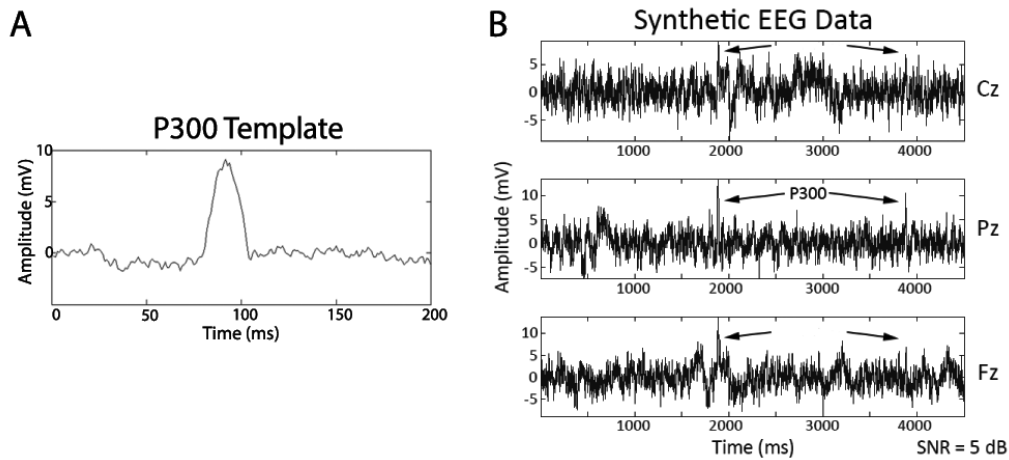}
\caption{A. The P300 template target signal.  B. An example of three channels (Cz, Pz, Fz) of synthetic ongoing EEG with two P300 target signal events (indicated by the arrows) at an SNR of 5~dB.}
\label{fig:EEG}
\end{figure}

Since the ``signal-plus-noise'' model (\ref{eq:bmi:signal-present}) reduces to the ``noise-only'' model (\ref{eq:bmi:signal-absent}) as $\alpha \to 0$, one would expect that the odds ratio should go to one, $\mbox{OR} \to 1$, as $\alpha \to 0$.  This can be accomplished by setting $\hat{\alpha}=0$ and letting $\sigma_\alpha \to 0$ in which case the truncated Gaussian prior for $\alpha$ (\ref{eq:BMI:truncated-prior-alpha}) collapses to a delta function. The odds ratio in (\ref{eq:bmi:odds-ratio-initial}) shows this limiting behavior as the argument of the exponential function approaches zero, and $Z_d / Z_\alpha \to 1$.  Another way to ignore the signal is to set $C_m = 0$, in which case $D=S^2$, $E=S^2\hat{\alpha}$. Again the argument of the exponential function in (\ref{eq:bmi:odds-ratio-initial}) vanishes and ratio of the normalizing constants approaches one.

To analyze the performance of the log OR filters, we generated synthetic electroencephalographic (EEG) data representing both the EEG background and the P300 evoked response, which is a brain response commonly used in BCI applications \cite{Sellers+etal:2006} (Figure \ref{fig:EEG}A).  Using the MATLAB code provided by Yeung,  Bogacz, et al. \cite{Yeung+Bogacz:EEG}, three channels of synthetic EEG data were generated to simulate recordings from scalp locations: Cz, Pz and Fz.  A current dipole model was used to scale the synthetic recordings from the different channels \cite{Knuth+Vaughan:99}.  The data from each of these channels consisted of 300 epochs each being 800 ms in length and comprised of 200 samples, which is consistent with a sampling rate of 250 Hz. Thirty epochs were selected to each host a single stereotypic P300 response at random latencies.  The remaining 270 epochs exhibited only ongoing background EEG (noise).

To study the effect of the Signal-to-Noise-Ratio (SNR) on the log OR filter performance, we
created 17 data sets where the SNR, calculated by the formula
\begin{equation}
\mbox{SNR}_{dB} = 10 \log_{10} \left( \frac{A_{signal}}{A_{noise}} \right)^2,
\end{equation}
was varied in integral steps from -6~dB to 7~dB as well as 10, 15 and 20~dB covering the typical SNR range seen in BCI and
EEG applications.  Figure \ref{fig:EEG}B illustrates synthetic ongoing EEG recordings with two target P300 signals (Figure \ref{fig:EEG}A) at an SNR of 5~dB.

\begin{figure}
\centering
\includegraphics[scale = 0.375]{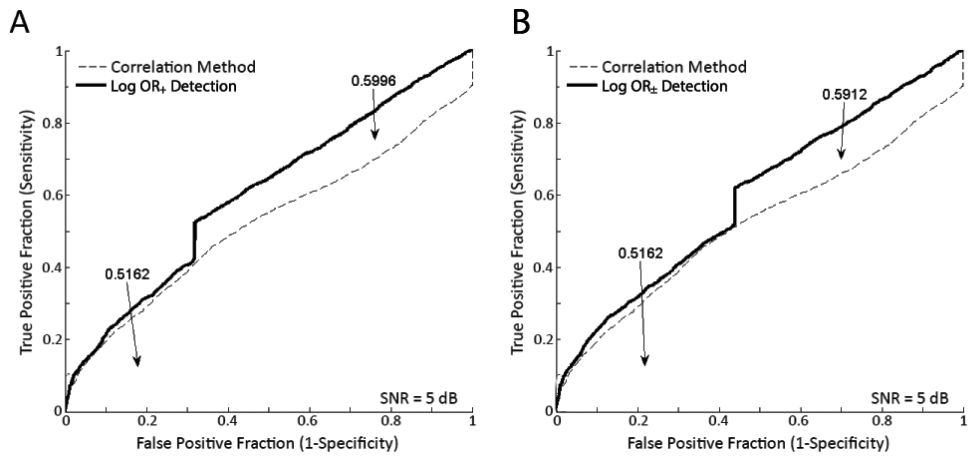}
\caption{A. This illustrates the ROC curves for both the Correlation Detection Method and the $\mbox{log OR}_{+}$ Detection Method in the case of SNR = 5~dB. Note that the $\mbox{log OR}_{+}$ Detection has a greater area under the curve (0.5996 as opposed to 0.5162 for Correlation), which indicates better performance over the Correlation Method. B. This figure illustrates the ROC curves for both the Correlation Detection Method and the $\mbox{log OR}_{\pm}$ Detection Method in the case of SNR = 5~dB.  While the $\mbox{log OR}_{\pm}$ Detection performs better than Correlation (0.5912 as opposed to 0.5162 for Correlation), it does not do quite as well as $\mbox{log OR}_{+}$ Detection in the case of SNR = 5~dB.}
\label{fig:ROC}
\end{figure}

\begin{figure}
\centering
\includegraphics[scale = 0.5]{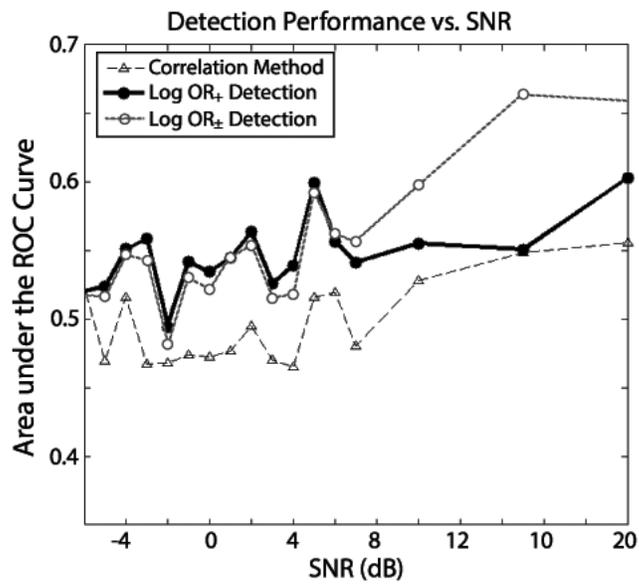}
\caption{A comparison of the performance of the two log OR methods, $\mbox{log OR}_{+}$ and $\mbox{log OR}_{\pm}$, and the Correlation Method as quantified by the areas under their respective ROC curves as a function of SNR.  The $\mbox{log OR}_{+}$ filter consistently performs better for low SNR, but is outperformed by $\mbox{log OR}_{\pm}$ Detection for SNR $>$ 5~dB. }
\label{fig:ROC-performance}
\end{figure}

The selection of a detection threshold value is a difficult task.  As the detection threshold increases, the sensitivity decreases
while the specificity increases, which means that the false positive fraction (1\hyp{}specificity) decreases. To study the performance of the log OR detection filter we compared it to the standard Correlation Method by producing Receiver Operating Characteristics (ROC) curves.  To do this we calculate sensitivity and (1\hyp{}specificity) for each distinct value of the detection measure (i.e. log OR / Correlation) to consider it as a candidate for detection cutoff. By plotting (1\hyp{}specificity) versus sensitivity, the efficacy of the detection method can be quantified by the area under the ROC curve \cite{Obuchowski:2005}.  Figure \ref{fig:ROC} compares the ROC curves of the Correlation Method with the $\mbox{log OR}_{+}$ Method (Figure \ref{fig:ROC}A) and the $\mbox{log OR}_{\pm}$ Method (Figure \ref{fig:ROC}B) obtained for target signals with an SNR of 5~dB.  These figures indicate that at this particular SNR, the log OR Detection filters outperform the traditional Correlation Method.  Figure \ref{fig:ROC-performance} provides a comparison of the performance of the two log OR methods, $\mbox{log OR}_{+}$ and $\mbox{log OR}_{\pm}$, and the Correlation Method as quantified by the areas under their respective ROC curves as a function of SNR for SNRs ranging from -6~dB to 20~dB.  The results indicate that the log OR detection filters based on Bayesian model testing consistently outperformed traditional Correlation.  Moreover, we see that the $\mbox{log OR}_{+}$ filter consistently performs better for low SNR, but is outperformed by $\mbox{log OR}_{\pm}$ Detection for SNR $>$ 5~dB.

\subsection{Light Sensor Characterization}
In this example, based on the work by Malakar, Gladkov and Knuth \cite{Malakar+etal:2013}, we demonstrate the use of Bayesian evidence to select the model order for a Gaussian mixture model of a light sensor, which was used in a robotics application \cite{Malakar+etal:2013}.  The problem involved identifying an accurate and efficient model of a LEGO light sensor (LEGO $\#$9844). The sensor consists of a photodiode-LED pair where the LED is used to illuminate the surface and the photodiode is used to measure the intensity of the reflected light.  The sensor integrates the light arriving from a spatially distributed region within its field of view, weighted by its spatial sensitivity function (SSF).  The goal was to model the SSF so that we could make accurate predictions of how the light sensor would respond when placed above a surface with a known albedo pattern.  We considered a mixture of Gaussians (MoG) model for the SSF in the sensor frame $(x',y') = (x-x_i, y-y_i)$,
\begin{multline}
SSF(x', y') = \\
\frac{1}{K} \sum_{n=1}^N{ a_n \exp{ \left[ \{ A_n(x'-u'_n)^2 + B_n(y'-v'_n)^2 +2 C_n (x'-u'_n)(y'-v'_n) \} \right] } }
\end{multline}
where $a_n$ and $(u'_n,v'_n)$ denote the amplitude and center of the $n^{th}$ Gaussian, respectively, where its covariance matrix elements are denoted by $A_n$, $B_n$ and $C_n$. The factor $K$ is the normalizing constant to ensure that the SSF integrates to unity in the case of a white surface.

\begin{figure}
\centering
\includegraphics[scale = 0.4]{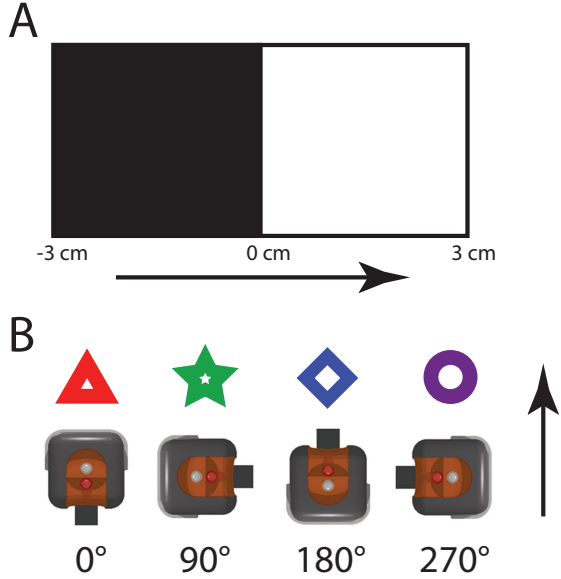}
\caption{A. An illustration of the black-and-white calibration surface. B. An illustration of the four orientations of the light sensor along with the colored symbols used to represent the intensities in Figure \ref{fig:SSF_data} recorded at each orientation with respect to the calibration surface as indicated by the black arrows.}
\label{fig:SSF_experiment}
\end{figure}

The MoG model is sufficiently general to be able to well-describe the SSF by varying the number of Gaussians. We considered four models consisting of one, two, three and four Gaussians. Each Gaussian in the mixture requires six parameters to be estimated
$\theta_n = {a_n, u_n, v_n, A_n, B_n, C_n}$,
where the subscript $n$ indexes the Gaussian in the mixture.

\begin{figure}
\centering
\includegraphics[scale = 1.0]{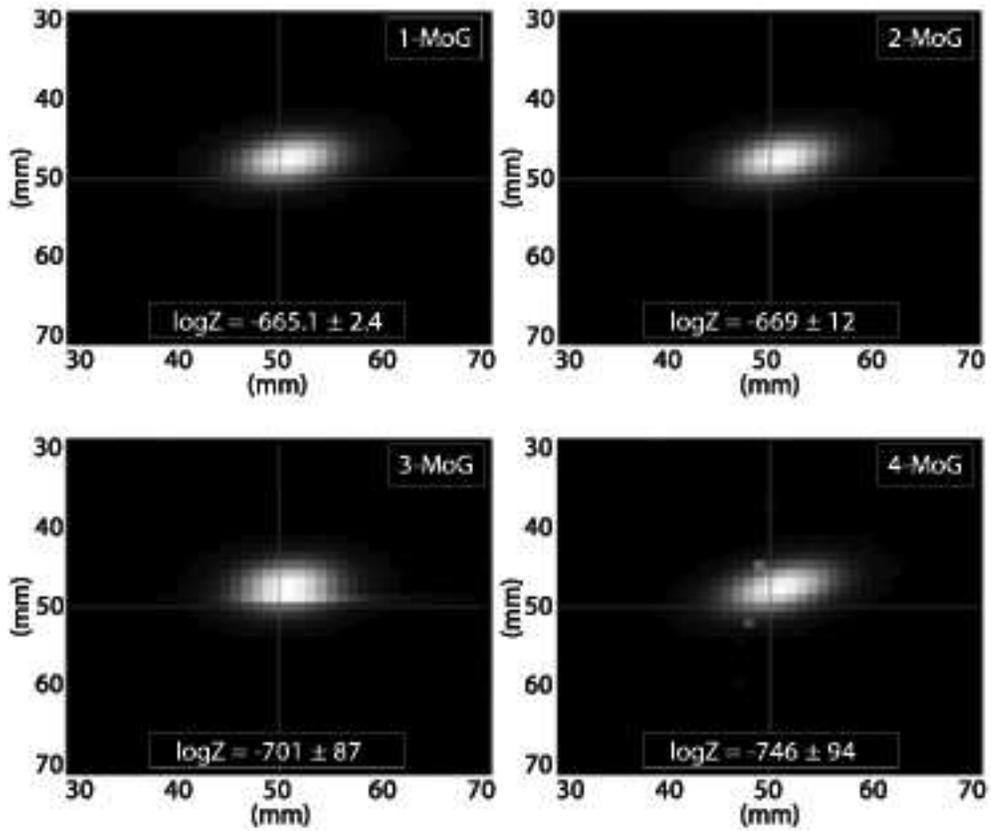}
\caption{An illustration of the four resulting MoG SSF Models along with their log-evidence values. The single Gaussian model (1-MoG) was found to have the greatest evidence, and therefore was selected as the optimal model of the light sensor SSF function.}
\label{fig:SSF_MoGs}
\end{figure}

\begin{table}[ht]
\centering
\begin{tabular}{l c c}
MoG Model & $\log Z$	&  $\#$ of Params \\
\hline
1 Gaussian  & $-665.1 \pm 2.4$ & $6$  \\
2 Gaussians & $-669 \pm 12$ & $12$ \\
3 Gaussians & $-701 \pm 87$ & $18$ \\
4 Gaussians & $-746 \pm 94$ & $24$ \\
\hline
\end{tabular}

\caption{This table lists the log-evidence ($\log Z$) values estimated for the MoG SSF models of various model orders.  The simplest model consisting of a single Gaussian (1-MoG) was found to be the most probable model.
However, the increasing uncertainty in the log-evidence estimates strongly suggests that this implementation of nested sampling is experiencing difficulties in handling the degeneracies in the mixture model.}
\label{tab:MoG_table}
\end{table}

In order to infer the model, we collected data by performing a series of experiments by recording intensities as the sensor was moved along a known surface (Figure  \ref{fig:SSF_experiment}A).  The sensor was held at a height of 14~mm above the surface in one of four orientations illustrated in Figure \ref{fig:SSF_experiment}B.  Intensities were recorded at increments of 1~mm steps as the sensor was moved in the direction of the arrow.  In addition to the surface illustrated in Figure  \ref{fig:SSF_experiment}A, we also presented the sensor with four corner patterns designed to break remaining symmetries.

\begin{figure}
\centering
\includegraphics[scale = 1.2]{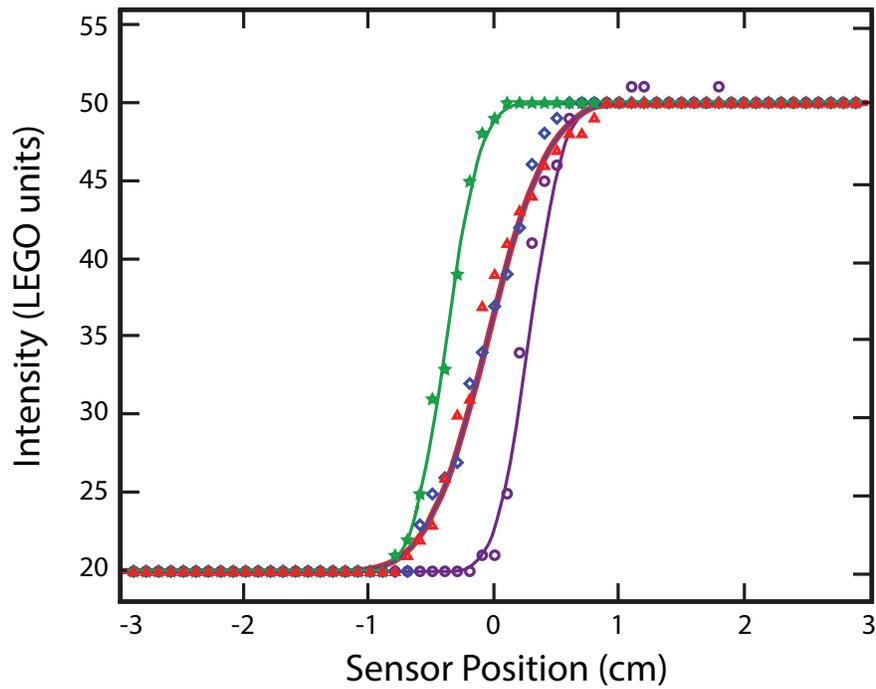}
\caption{The 1-MoG SSF model, with the maximum log-evidence, is used to predict the sensor intensity (solid curves) when applied to the black-and-white calibration surface and compare it with the recorded intensities (discrete symbols). The four symbols denote the four orientations of the sensor as indicated in Figure \ref{fig:SSF_experiment}. Note that since the 1-MoG SSF model is aligned with the measurement axes, the predicted curves for the $0^\circ$ orientation (red triangles) and $180^\circ$ orientation (blue diamonds) are identical and overlaid on top of one another in the center of the figure.}
\label{fig:SSF_data}
\end{figure}

Bayesian estimation of the MoG model parameters was performed using nested sampling with 300 samples,
and was repeated 20 times for each model order to obtain uncertainty estimates of the log-evidence.
We assigned uniform priors to the model parameters as well as a Student-t distribution to the likelihood.  The nested sampling algorithm was iterated while monitoring the log-evidence, and was stopped when the change in the consecutive log-evidence values was less than 1e-8.

Figure \ref{fig:SSF_MoGs} shows the four resulting MoG models of the SSF function.  Table \ref{tab:MoG_table} shows the evidence values for the competing models.  The 1-MoG model consisting of a single Gaussian had the greatest
mean log-evidence of -665.1, which is why it was selected as the optimal model.
Figure \ref{fig:SSF_data} compares the predictions (black) made by the 1-MoG SSF model to the observed intensities (red) showing excellent agreement.  The resulting SSF model obtained by maximizing the log-evidence was found to be both accurate and efficient, and was selected for use in further studies involving that light sensor \cite{Malakar+etal:2013}.

However, given the uncertainties of the log-evidence values of the models in Table \ref{tab:MoG_table}, selection of a \textit{\emph{best model}} based on the log-evidence alone is not clear.  These results suggest that this implementation of nested sampling is experiencing difficulties dealing with the multiple optima resulting from the degeneracies of the mixture model.  This is similar to the challenge faced by Chib using Gibbs sampling \cite{Chib:1995:gibbs}, as noted, and solved, by Berkhof et al. \cite{Berkhof+etal:2003}.  In the Conclusion, we will make some additional comments on selecting an optimal model order based on log-evidence estimates.
%, which was 6.4 greater than the 3-MoG model which had the next highest log-evidence of -671.9.  This means that the 1-MoG model is $\exp(6.4) \approx 600$ times more probable than the 3-MoG model.  For this reason, the optimal model is the one consisting of a single Gaussian (1-MoG Model).

%One might expect that the log-evidence should increase as the model order increases up until the point that the optimal model is reached and then decrease steadily as the model order increases further.  This is because the log-evidence results from interplay between the prior probability, which monotonically decreases as the model order increases, and the likelihood, which monotonically increases as the model order increases.  In the case where the simplest model has the greatest evidence, one might expect that the log-evidence will monotonically decrease as model order increases.  However, in this example, this is not the case since the 3-MoG model has a slightly greater log-evidence than the 2-MoG model.  This can happen since even though the prior is decreasing monotonically and the likelihood is increasing monotonically the rates at which these two competing factors change can result in non-monotonic log-evidence variations, which we see here.  Basically, the relative increase in the likelihood with respect to the 2-MoG model was greater than the relative decrease in the prior with respect to the 2-MoG model, resulting in a slight increase in log-evidence.

\subsection{Exoplanet Detection}

Our third example concerns the determination of the importance of various photometric effects in an exoplanetary system.  The details of this study by Placek, Knuth, et al. can be found in the following references \cite{Knuth+Placek+Richards:2012}\cite{Placek+Knuth+Angerhausen:EXONEST}.  Currently, the primary method of detecting and characterizing exoplanets involves the analysis of the time series resulting from the observations of unresolved light coming from a planetary system.  The presence of exoplanets around distant stars is known to produce at least four physical mechanisms that affect the observed photometric signal in very specific ways.  The first two effects originate from the planet itself.  As the planet orbits it's host star, it undergoes phases just as Venus and Mercury do in this Solar System from the perspective of Earth.  This will cause photometric variations since the amount of reflected light off of the atmosphere or surface of the planet will change throughout the planet's orbit.  By modeling the reflectance as Lambertian, one can model these stellar-normalized flux variations as
\begin{equation}
\frac{F_{R}(t)}{F_\star} = \frac{A_g}{2} \frac{{R_p}^2}{r(t)^2} \left(1 + \cos \theta(t) \right).
\end{equation}
where $A_g$ is the geometric albedo of the planet, which represents how effective the planet is at reflecting incident light back into space, $R_p$ is the planetary radius, $r(t)$ is the planet-star separation distance, $\theta(t)$ is the angle between the observer's line-of-sight and the line connecting the star to the planet, and $F_\star$ is the stellar flux.
Similarly, planets have a temperature and therefore emit thermal radiation.  This also contributes to the observed photometric signal and can be modeled for both day and night sides as
\begin{equation}
\frac{F_{T,d}(t)}{F_\star} = \frac{1}{2}(1 + \cos\theta(t)) \left( \frac{R_p}{R_\star} \right)^2 \frac{ \int B(T_d)K(\lambda) \, d\lambda}{\int B(T_{eff}) K(\lambda) \, d\lambda}
\end{equation}
where $R_\star$ and $T_{eff}$ are the stellar radius and effective temperature, respectively, $B(T)$ is the spectral radiance of a blackbody, and $K(\lambda)$ is the instrument response as a function of wavelength $\lambda$.  The expected stellar-normalized flux from the night-side $\frac{F_{T,n}(t)}{F_\star}$ is found using the night-side temperature of the planet $T_n$.

The remaining two effects are induced by the planet but involve the host star.  Stars and planets both orbit the center of mass of the system.  As the star revolves around the center of mass, an observer moving relative to that star will observe increases in the amount of flux emitted from the star as it approaches, and a decrease in flux as it recedes.  This is known as Doppler beaming and is a relativistic effect.  In the non-relativistic limit, the flux variations can be approximated as
\begin{equation}
\frac{F_{B}(t)}{F_\star} =  1 + 4\beta_r(t)
\end{equation}
where $\beta_r(t)$ is the component of stellar velocity along the line-of-sight.  This effect has the same frequency as the previous two, however the signal is shifted in phase by $\pi/2$.
Finally, due to the proximity of the planet to the star, the planet will induce tides on the stellar surface causing the star to appear as a prolate spheroid.  These tides will follow the planet in its orbit and result in flux variations at twice the orbital frequency since the cross-section of the star is changing throughout the orbit.  This effect is approximated by
\begin{equation}
\frac{F_{ellip}(t)}{F_\star} = \beta \frac{M_p}{M_\star} \left( \frac{R_\star}{r(t)} \right)^3 [\cos^2(\omega+\nu(t)) + \sin^2(\omega+\nu(t))\cos^2i]
\end{equation}
where $\beta$ is the gravity darkening exponent, $M_p$ and $M_\star$ are the planetary and stellar masses, respectively, $\omega$ is the argument of periastron, $\nu(t)$ is the true anomaly, and $i$ is the orbital inclination.

In order to obtain a predictive model for the total observed signal, one needs to sum the photometric contributions from each effect
\begin{equation}
F_{pred}(t) = F_\star \left(1 + \frac{F_p(t)}{F_\star} + \frac{F_{boost}(t)}{F_\star} + \frac{F_{ellip}(t)}{F_\star} + \frac{F_{Th,d}(t)}{F_\star} + \frac{F_{Th,n}(t)}{F_\star}\right).
\end{equation}

\begin{figure}
\centering
\includegraphics[scale = 0.24]{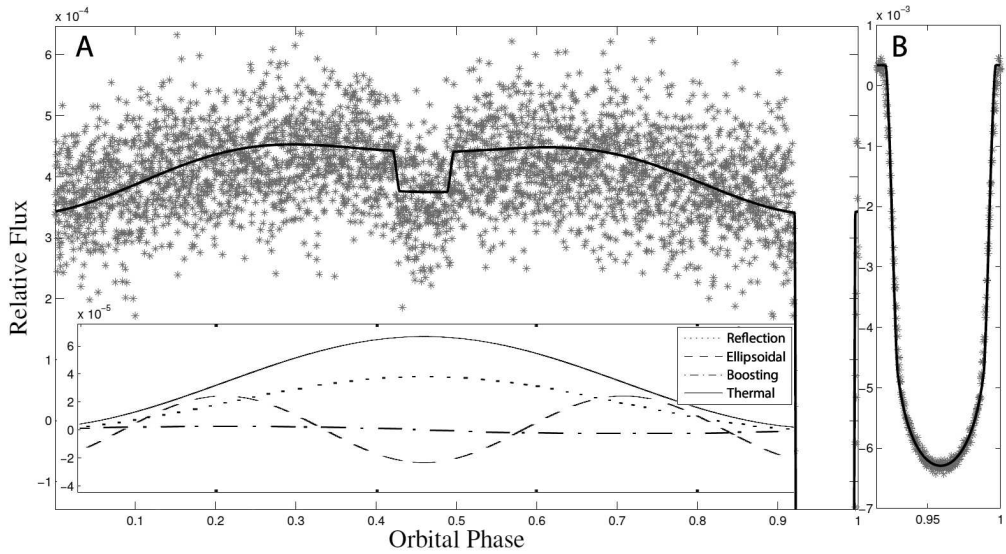}
\caption{A. A model fit (solid curve) to exoplanet KOI-13b data (asterisks) from the Kepler Space Telescope.  The secondary eclipse from the planet passing behind the star is centered in the plot between the phases 0.4 and 0.5.  The primary transit from the planet passing in front of the star occurs at the far right between phases 0.9 and 1.0.  The inset at the bottom shows the estimated photometric flux contributions from reflected light, ellipsoidal variations, Doppler boosting and thermal emissions.  B. A detailed illustration of the model fit to the primary transit.}
\label{fig:photometriceffects}
\end{figure}

Bayesian model selection allows one to effectively characterize exoplanetary systems.  Each of these four effects can be present in the data to varying degrees, or completely absent.  Thus, one can create a suite of models each comprised of a different subset of the four photometric effects.  Since all four effects depend on the orbital orientation of the planet, model testing also allows one to test between circular and eccentric orbits.  By calculating the evidence for each model, one could determine whether or not each effect is present in the data, and how large of a role each effect plays in describing the observed data.

As an example, we performed such model testing on data obtained from the Kepler Space Telescope for a confirmed exoplanet called KOI-13b.  KOI-13b is known as a short-period hot Jupiter since it orbits its host star in just $1.7637$ days and has a temperature of over $3500$ K.  This sort of exoplanet is expected to induce large ellipsoidal variations on its host star, produce significant thermal emission and less reflection.  This is due to the fact that most of the reflective condensates in the atmosphere, such as water and ammonium, are essentially burned off, significantly decreasing the planetary albedo.  A set of $18$ models were applied to KOI-13b (shown in Table \ref{tab:KOI13b_evidences}) and log-evidences were calculated for each one using the MultiNest algorithm \cite{Feroz+Hobson:2008}\cite{Feroz+etal:2009}\cite{Feroz:2013}, which is one of several inference engines included in our EXONEST Exoplanetary Explorer software suite \cite{Placek+Knuth+Angerhausen:EXONEST}.

In general, the noise is expected to be Gaussian-distributed about the mean exoplanetary signal.  Therefore, a Gaussian log-likelihood of the form
\begin{equation} \label{eq:exoplanet-logL}
\log L = -\frac{1}{2}\chi^2 - \frac{1}{2}N\log(2\pi\sigma^2)
\end{equation}
was used in each of the $18$ simulations, where $\sigma^2$ is the noise variance, $N$ is the number of datapoints, and $\chi^2$ is the sum of the squared residuals divided by $\sigma^2$.
The noise variance was treated as a free parameter to be estimated by MultiNest. Often, stars display short-period variability induced by oscillations, starpots, and other effects. This variability can lead to correlated (red) noise, which can deviate from a Gaussian distribution. In that case, one may adopt a more detailed likelihood function that utilizes a nearest-neighbor approach to deal with noise correlations in the time series signal \cite{Sivia&Skilling}.

%\begin{table}[ht]
%\centering
%\begin{tabular}{l c c c c}
%			Model		&  Circular				&  Eccentric 				&  $\chi^2$ (ppm)	& Model Parameters \\
%\hline
%R						& $37\,108.0 \pm 0.4$	& $37\;659.0 \pm 5.4 $ 	& $3.7493$ 		& $7$\\
%
%B						& $36\,970.0 \pm 4.0$     & $37\,166.0 \pm 1.9 $	&$4.7056$		& $7$\\
%
%E						& $36\,555.0 \pm 0.5  $    & $37\,581.0 \pm 0.4 $	&$6.7226$		& $7$ \\			
%
%R+B						& $37\,108.0 \pm 0.5 $ 	& $37\,670.0 \pm 2.9 $ 	&$3.7391$		& $8$\\
%
%R+E						& $37\,701.0 \pm 0.5 $	& $37\,704.0 \pm 2.7$ 	&$3.7257$		&$8$\\
%
%B+E 						& $36\,577.0 \pm 0.8  $	& $37\,634.0 \pm 2.8 $  	&$6.5497$		&$7$ \\
%
%R+B+E					& $37\,703.0 \pm 1.1 $	& $37\,748.0 \pm 1.1 $ 	&$3.4503$		&$8$\\
%
%T+B+E 					& $37\,703.0 \pm 1.1 $ 	& $\bf{37\,764.0 \pm 8.3} $ &$3.3674$		&$9$ \\
%
%R+B+E+T 				& ... 					& $\bf{37\,765.0 \pm 0.9} $ &$3.3686$		&$10$\\
%
%\hline
% Null & \multicolumn{2}{c}{$36\,143.0 \pm 1.0$} & & $2$\\
%\end{tabular}

\begin{table}[ht]
\centering
\begin{tabular}{l c c c c}
			Model		&  Circular				&  Eccentric 				&  $\chi^2$	& Model Parameters \\
\hline
R						& $37\,108.0 \pm 0.4$	& $37\;659.0 \pm 5.4 $ 	& $2023$ 		& $7$\\

B						& $36\,970.0 \pm 4.0$     & $37\,166.0 \pm 1.9 $	&$2539$		& $7$\\

E						& $36\,555.0 \pm 0.5  $    & $37\,581.0 \pm 0.4 $	&$3627$		& $7$ \\			

R+B						& $37\,108.0 \pm 0.5 $ 	& $37\,670.0 \pm 2.9 $ 	&$2018$		& $8$\\

R+E						& $37\,701.0 \pm 0.5 $	& $37\,704.0 \pm 2.7$ 	&$2010$		&$8$\\

B+E 						& $36\,577.0 \pm 0.8  $	& $37\,634.0 \pm 2.8 $  	&$3534$		&$7$ \\

R+B+E					& $37\,703.0 \pm 1.1 $	& $37\,748.0 \pm 1.1 $ 	&$1862$		&$8$\\

T+B+E 					& $37\,703.0 \pm 1.1 $ 	& $\bf{37\,764.0 \pm 8.3} $ &$1817$		&$9$ \\

R+B+E+T 				& ... 					& $\bf{37\,765.0 \pm 0.9} $ &$1818$		&$10$\\

\hline
 Null & \multicolumn{2}{c}{$36\,143.0 \pm 1.0$} & & $2$\\
\end{tabular}

\caption{MultiNest log-evidences for 18 different models applied to the photometric signal of KOI-13b.  Each model is named after the effects that it takes into account (Reflection - R, Doppler Beaming - B, Ellipsoidal Variations - E, Thermal Emissions - T). The models most favored to describe the data are in bold. Note that with the reflectance and thermal emissions models used in this study, reflected light intensity and thermal emissions cannot be distinguished in a circular orbit.  For this reason, that specific case was not analyzed. The $\chi^2$ values for the best fit eccentric models and the number of parameters for each are also listed. The two most probable models correspond to the best fit models according to the $\chi^2$ criterion. The Null Planet model consisted of two model parameters: the noise level $\sigma \in [10^{-6}, 10^{-4}]$ and the baseline flux $\in [-0.1, 0.1]$. Last, note that the log-evidence values presented are positive due to the fact that the noise variance $\sigma^2$ is very small for this data with $\sigma$ values being as low as $10^{-6}$. This results in a large positive value for the second term of the log likelihood (\ref{eq:exoplanet-logL}), which dominates the evidence integral.}
\label{tab:KOI13b_evidences}
\end{table}

Each model was applied twice for circular and eccentric orbits.  The simpler models are shown at the top of Table \ref{tab:KOI13b_evidences} and they increase in complexity moving down the table.  The two models most favored to describe the data are those including thermal emission, Doppler boosting, and ellipsoidal variations ($\log Z = 37\,764 \pm 8.3$), and reflection, thermal emission, Doppler boosting, and ellipsoidal variations ($\log Z = 37\,765.0 \pm 0.9$), which is illustrated in Figure \ref{fig:photometriceffects}.  Based on the uncertainties on the log-evidences, these two models have an essentially equal probability to describe the observed data.    This also means that adding the reflection effect to thermal emissions does not yield a significantly better fit as indicated by the $\chi^2$ values, which indicate a difference in the sum of the squared residuals of only $0.12\%$.  This is to be expected for planets similar to KOI-13b since they have very low albedos and are very hot due to the proximity to the host star. In each case, the eccentric model is more favored than the circular.

The astute reader will note that the log-evidence values in Table \ref{tab:KOI13b_evidences} are positive, which indicates that there are large positive log likelihood values in the integral.  Since the likelihood, and hence the evidence, are density functions and have units, this is a result of the choice of units for flux.  The log likelihood (\ref{eq:exoplanet-logL}) is the sum of two terms.  The first term is unit-less, whereas the second term depends on $\log(\sigma)$, which can change signs depending on the units.\footnote{See \url{http://blog.stata.com/2011/02/16/positive-log-likelihood-values-happen/} for more information on this effect.} As a check, these can be estimated by considering the Null Planet model with zero baseline and a noise level of $\sigma = 5 \times 10^{-5}$.  In this case, one can use the fact that there are $N = 4187$ flux data points with a standard deviation of $4.3050 \times 10^{-5}$ resulting in
% $\chi^2 = 7.7600 \times 10^{-6}$
\red $\chi^2 = 3104$
and a log likelihood (\ref{eq:exoplanet-logL}) of
$36066$ \black
%$36828$
for those particular model parameter values.  Since this is the logarithm of the likelihood, large positive values like this dominate the evidence integral resulting in a log-evidence of $log Z = 36143.0 \pm 1.0$ for the Null Planet model.  Since there is a planet present, this represents a lower bound to the positive log-evidence values obtained in Table \ref{tab:KOI13b_evidences}.

By comparing the results from multiple models some important facts about the KOI-13b are revealed.  First, both the $\chi^2$ values and the log-evidences indicate that reflected light (or thermal emissions, which are similar to reflected light) is a prominent component in the photometric signature.  One can also note that the R+E model, which describes reflection and ellipsoidal variations ($\log Z = 37\,704 \pm 2.7$), is only marginally less probable than the models that include all three photometric effects.  This further implies that ellipsoidal variations also play a significant role in the observed data.  This indicates an additional advantage to comparing and contrasting sets of models based on the $\chi^2$ values and the log-evidences. By calculating the Bayesian evidence and incorporating model testing by turning on and off certain photometric effects, one can effectively characterize planetary systems, as well as use it as a planetary confirmation procedure.

\subsection{Force field selection in biomolecular structure determination}
Last, we present an example from Habeck in structural biology \cite{Habeck:2011}. NMR spectroscopy allows us to determine the three-dimensional structure of complex biomolecules such as proteins at atomic resolution. However, often the data are not sufficient to determine the structure without additional guidance from molecular mechanics force fields. These force fields can be very complex, which slows done the structure calculation. Therefore, the force fields used in biomolecular structure determination typically neglect important contributions such as electrostatic or solvent interactions and rather work with a minimalist force field. On the other hand it is clear that by choosing more realistic force fields the results obtained from challenging data will be more useful.

Current practice to calculate biomolecular structures is to set up a cost function (the so-called hybrid energy) $\lambda D(x,d) + E(x)$ that is comprised of a data fitting term $D(x,d)$ weighted by $\lambda$ and a force force field $E(x)$ where $x$ are the conformational degrees of freedom of the biomolecule (e.g. the Cartesian coordinates of all atoms or dihedral angles) and $d$ represents relevant data. Inferential structure determination  (ISD) \cite{Rieping:2005} is a strictly probabilistic approach to solve structure determination problems. It not only allows us to estimate the appropriate weight of the data $\lambda$ \cite{Habeck+etal:2006}, but also to compare two alternative force fields in the light of given experimental data \cite{Habeck:2011} as well as determine the best weight of the force field \cite{Mechelke:2012}. ISD models the data $d$ probabilistically such that
\begin{equation}
P(d|x,M,I) = \frac{1}{Z_D(\lambda,d)} e^{-\lambda D(x,d)}
\end{equation}
where $Z_D(\lambda,d)$ is a normalizing constant that depends on the chosen model $M$ to assess discrepancies between observed data $d$ and predictions made by the forward  model. The force field $E(x)$ is incorporated using a Boltzmann distribution as prior probability over the conformational degrees of freedom:
\begin{equation}\label{eq:boltzmann}
P(x|M,I) = \frac{1}{Z_E(\beta)} e^{-\beta E(x)}
\end{equation}
where $Z_E(\beta)$ is the partition function of the Boltzmann distribution and normalizes the prior. In the most general case the inverse temperature $\beta$ of the force field is unknown because, as explained above, we cannot afford to work with realistic force fields but have to make drastic simplifications. Therefore also the ``temperature'' of the minimalist force field is no longer identical to the temperature at which the experiments were carried out, but is instead an unknown hyperparameter \cite{Mechelke:2012}.

Here, we compare two different force fields that are used in biomolecular modeling. Both aim to describe van der Waals interactions between atoms that are not linked via a covalent bond. The first is a quartic repulsion term that drops to zero when the distance between two atoms $r_{ij}$ is larger than the sum of their van der Waals radii $R_i$  \citep{Habeck:2005}:
\begin{equation}\label{eq:quartic}
E_\mathrm{quartic}(r_{ij}) = \left\{\begin{array}{c c} (r_{ij} - R_i - R_j)^4; & r_{ij} \le R_i + R_j\\
0; & r_{ij} > R_i + R_j\end{array} \right.
\end{equation}
This force field ignores the attractive contribution of the van der Waals interaction. An alternative force field that takes the attractive term into account is used in the Rosetta software \citep{Kuhlman:2003}. This is a Lennard-Jones potential that is linearly ramped to finite values as $r_{ij}$ approaches zero and vanishes for distances larger than a cutoff distance of $R_{\mathrm{cut}} = 5.5$ \AA. The potential function is:
\begin{equation}\label{eq:LJ}
E_\mathrm{LJ}(r_{ij}) = \left(\frac{R_i + R_{j}}{r_{ij}}\right)^{12} - 2 \left(\frac{R_i + R_{j}}{r_{ij}}\right)^6
\end{equation}
for
$0.6 (R_i + R_j) < r_{ij} \leq 5 \, \mathrm{\AA}$
and continues linearly to the left and right of this interval.
%
%\begin{equation}
%E_\mathrm{LJ}(r_{ij}) = \left\{\begin{array}{c c}
%
%(\frac{R_i + R_{j}}{r_{ij}})^{12} - 2 (\frac{R_i + R_{j}}{r_{ij}})^6 ; & r_{ij} \le \mathrm{sw} (R_i + R_j) \\
%
%(\frac{R_i + R_{j}}{r_{ij}})^{12} - 2 (\frac{R_i + R_{j}}{r_{ij}})^6 ; & \mathrm{sw} (R_i + R_j) < r_{ij} \le R_{\mathrm{rlin}} \\
%
%(2-a) a (r_{ij} - R_{\mathrm{cut}}); & R_{\mathrm{rlin}} < r_{ij} \le R_{\mathrm{cut}} \\
%
%0; & r_{ij} > R_{\mathrm{cut}}\end{array} \right.
%\end{equation}
%where $a=(R_i + R_j)^6 / R_{\mathrm{rlin}}^6$, the switch $\mathrm{sw} = 0.6$.
%    a = 1 / ROSETTA_SWITCH;
%    a*= a * a;
 %   a*= a;
%    d *= -12 * (a - 1) * a / (ROSETTA_SWITCH * d0);
%    return k * (d + (13 * a - 14) * a);
%

We compare these two force fields in the light of NMR data measured on the Fyn-SH3 domain, a small signaling domain of 59 amino acids in length. The data are sparse and comprised of 154 inter-proton distances measured on a deuterated sample \citep{Rieping:2005}. We ran a parallel tempering simulation for each of the two force fields. The parallel tempering schedule is two-dimensional \citep{Habeck:2005b}: The first replica parameter $\lambda$ is the inverse temperature and gradually switches off the data, whereas the second parameter is Tsallis' $q$ used to deviate from the Boltzmann ensemble (\ref{eq:boltzmann}). The Tsallis ensemble approaches the Boltzmann ensemble for $q\to 1$ and is used here only for convenience because neighboring replicas will show a higher overlap due to the fatter tails of the Tsallis ensemble. Fifty replicas were set up in which $\lambda$ varied from 0.1 to 1.0 and $q$ varied from 1.06 to 1.0; we used the same combination of $(\lambda,q)$ values for both force fields.

\begin{figure}
\centering
\resizebox{\textwidth}{!}{\includegraphics{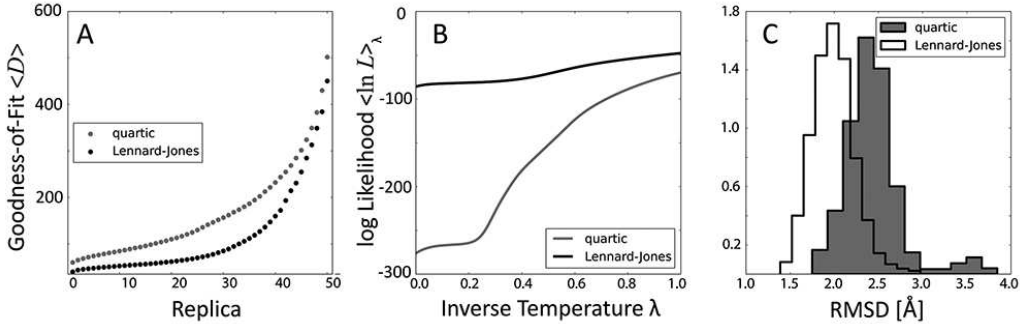}}
\caption{Comparison of force fields in biomolecular structure determination. A: Average goodness of fit $\left\langle D \right\rangle_{\lambda,q}$ in the 50 replicas at varying $\lambda$ and $q$ used to sample the posterior. B: Average log likelihood $\left\langle \log L \right\rangle_\lambda$ obtained with the estimated DOS. C: Accuracy as measured by the root mean square deviation (RMSD) of the structures sampled from the posterior with the crystal structure.}\label{fig:sh3}
\end{figure}

Figure \ref{fig:sh3}A shows the average goodness of fit $\left\langle D \right\rangle_{\lambda,q}$ (negative log likelihood) for each of the 50 replicas. It is already apparent from this figure that the Lennard-Jones potential (\ref{eq:LJ}) results in a better goodness of fit than the purely repulsive potential [Eq. (\ref{eq:quartic})]. We applied histogram re-weighting to estimate the density of states (DOS) from the replica simulations \citep{Habeck:2012,Habeck:2012b}. The estimated DOS can be used to calculate the expected log-likelihood as a function of the inverse temperature $\left\langle \log L \right\rangle_\lambda$ and  apply thermodynamic integration, which would not be possible without the help of the DOS because $(\lambda,q)$ are varied simultaneously. Figure \ref{fig:sh3}B shows the expected  log likelihood $\left\langle \log L \right\rangle_\lambda$ as a function of the inverse temperature, i.e. the integrand of thermodynamic integration equation (\ref{eq:thermodynamicintegration}). Alternatively, we can evaluate the partition function (\ref{eq:partitionfunction2}) to compute the evidence. Both approaches are equivalent and give the same result. The evidence clearly favors the Lennard-Jones potential ($\log Z = -69$) over the potential based on a quartic repulsion term ($\log Z = -166$). The Lennard-Jones potential is not only more supported by the NMR data but also results in a more accurate structure ensemble. The root mean square deviation (RMSD) between members of the posterior ensemble and the crystal structure, serving here as a reference, is systematically shifted towards better values when using the Lennard-Jones potential (see Fig. \ref{fig:sh3}C).

\section{Conclusion}

In this paper we have reviewed the concept of the Bayesian evidence (marginal likelihood) and the related concepts of Bayes factors and odds ratios, which quantify the probability of one model over another based on the selected models and the data.  That is, the degree to which the data implies a given model.  In addition to discussing the analytic treatment of the foundations, we have focused mainly on approximate and numerical techniques such as the Laplace approximation, variational Bayes, thermodynamic integration and stochastic integration via Monte Carlo methods.

The discussions regarding these methods were supplemented by four examples with the intention of demonstrating Bayesian model testing in different scientific domains: signal detection (BCI) \cite{Mubeen+Knuth:2014}, sensor characterization (robotics) \cite{Malakar+etal:2013}, scientific model selection (exoplanet characterization) \cite{Knuth+Placek+Richards:2012}\cite{Placek+Knuth+Angerhausen:EXONEST} and molecular force characterization (structural biology) \cite{Habeck:2011}. Together these applications demonstrate the power of Bayesian model testing in a variety of contexts leading to improved signal processing algorithms, improved instrument models, as well as a deeper understanding of physical systems at scales ranging from the astronomic to the microscopic.  These examples, which involve detailed theoretical signal models, do not begin to cover the vast array of inference problems and underlying models that one could consider.  For example, nonparametric models find great use in domains where detailed signal models are lacking. Examples of such models include Gaussian Processes \cite{Gibbs:1998}\cite{MacKay:1998}\cite{Rasmussen+Williams:2006}\cite{Osborne:2010} and generalized autoregressive models \cite{Roberts+Penny:2002:vb}\cite{Penny+Kiebel+Friston:2006:vb}.  As demonstrated in the provided references, Bayesian model testing works well with those nonparametric models as well.

In the examples given in this paper, model selection was based on the evidence or Bayes factors alone.  However, it is important to remember that probability theory is not decision theory \cite{Berger:1985}.  That is, there are other factors involved in any decision-making process that can be described by a utility function that maximizes expected utility or minimizes expected loss.  For this reason, it is strongly recommended that model selection be performed by considering both the probability of a model and the expected utility function \cite{Hogg:decisions}.  In practice, this can quite challenging as it is often difficult to identify and to quantify such utility especially in situations where there are multiple factors involved.  This remains an active area of research.

%It should be mentioned that there are other approximate methods for performing model testing. One such method is variational Bayes which aims to approximate the posterior with a simpler function that can be integrated to obtain a lower bounds for the Bayesian evidence.  Another popular technique is the more easily computed Akaike Information Criterion (AIC) \cite{Akaike:1974}, which is based on the Kullback-Leibler divergence.  The AIC has been shown to be an approximation to the Bayesian Odds Ratio; however, as such, it lacks important terms \cite{Zellner:1978}.  Last, the Bayesian Information Criterion (BIC) ***
%
%***

\section{Acknowledgements}
This paper was written for a special issue dedicated to William J. Fitzgerald.  While only once did I (Knuth) have the pleasure to meet Prof. Fitzgerald, I was influenced by him indirectly through several of his students, Ali Taylan Cemgil, Ercan E. Kuruoglu, and Robin D. Morris, who in addition to always speaking very highly of him were each clearly excellently trained as creative, talented and responsible members of the scientific community.  This research was supported in part by a University at Albany Faculty Research Awards Program (FRAP-A) Award (Knuth), a Deutsche Forschungsgemeinschaft (DFG) grant HA 5918/1-1 (Habeck) as well as two University at Albany Benevolent Research Grant Awards (Malakar and Mubeen).  Additional N. Malakar support carried out at the Jet Propulsion Laboratory, California Institute of Technology, under a contract with the National Aeronautics and Space Administration.  The authors would like to thank the anonymous reviewers for their careful, thorough, and extensive comments, which have served to improve the quality of this manuscript.

\section{Vitae}
Kevin H. Knuth is an Associate Professor in the Departments of Physics and Informatics at the University at Albany, Albany NY USA. He is Editor-in-Chief of the journal Entropy, and is the co-founder and President of a robotics company, Autonomous Exploration Inc, and a former NASA research scientist. He has 20 years of experience in applying Bayesian and maximum entropy methods to the design of machine learning algorithms for data analysis applied to astronomy and the physical sciences. His current research interests include the foundations of physics, autonomous robotics, and searching for and characterizing extrasolar planets.\\
\\
Michael Habeck received his Ph.D. in Biophysics in 2004 from the University of Regensburg, Germany. In 2009, he started an independent research group at the Max Planck Institute for Developmental Biology in T{\"u}bingen, Germany. Since 2013 he is a group leader in Computational Structural Biology at the University of G{\"o}ttingen and the Max Planck Institute for Biophysical Chemistry. His major research interests are in the application of Bayesian inference to data analysis problems arising in structural biology.\\
\\
Nabin K. Malakar received his Ph.D. degree in physics from the University at Albany (SUNY), Albany NY USA, in 2011.  He is currently a postdoctoral scholar at the Jet Propulsion Laboratory, California Institute of Technology, in Pasadena. He is working on the development, validation and evaluation of new land surface temperature and emissivity products from the MODIS instrument onboard the Terra and Aqua satellites. His research interests includes identification of relevant variables in a physical phenomena to improve our understanding of atmospheric processes, as well as algorithm development for various applications of remote sensing data.\\
\\
Asim M. Mubeen received his M.Sc. degree in physics from the Punjab University Lahore, Pakistan in 1998, and M.S. degree in physics from the University at Albany (SUNY), Albany NY USA, in 2007.  He is currently a Ph.D student in Department of Physics at University at Albany (SUNY), Albany USA.  Since December 2013, he is an Assistant Research Scientist, in the Geriatrics Division at Nathan Kline Institute, Orangeburg, NY.  His research interests include Bayesian inference, digital signal processing, brain computer interface, image processing, and diffusion tensor imaging (DTI).\\
\\
Ben Placek received his  Ph.D. in physics from the University at Albany (SUNY), Albany NY USA, in 2014 and is currently a Physics Instructor at Schenectady County Community College. His research is focused on exoplanet detection and characterization, and he has worked to develop the EXONEST algorithm, which employs Bayesian methods to improve exoplanet characterization.

\section{References}
%% The Appendices part is started with the command \appendix;
%% appendix sections are then done as normal sections
%% \appendix

%% \section{}
%% \label{}

%% If you have bibdatabase file and want bibtex to generate the
%% bibitems, please use
%%
\bibliographystyle{elsarticle-num}
\bibliography{C:/Users/KK952431/kevin/files/papers/bibliography/knuth}

%% else use the following coding to input the bibitems directly in the
%% TeX file.

%\begin{thebibliography}{00}
%
%%% \bibitem{label}
%%% Text of bibliographic item
%
%\bibitem{}

%\end{thebibliography}
\end{document}